\documentclass[a4paper,english,aps, floatfix]{revtex4}
\usepackage{mathptmx}
\usepackage[T1]{fontenc}
\usepackage[latin9]{inputenc}
\usepackage{graphicx}
\usepackage{esint}

\providecommand{\tabularnewline}{\\}

\usepackage{babel}

\begin{document}

\title{Anomalous resonant production of the fourth family up type quarks
at the LHC}

\author{\.{I}. T. Çak\i{}r}

\email{tcakir@mail.cern.ch}

\affiliation{Department of Physics, CERN, 1211 Geneva 23, Switzerland}

\author{H. Duran Y\i{}ld\i{}z}

\email{hyildiz@mail.cern.ch}

\affiliation{Department of Physics, Dumlup\i{}nar University, Faculty of Arts
and Sciences, Kütahya, Turkey.}

\author{O. Çak\i{}r}

\email{ocakir@mail.cern.ch}

\affiliation{Department of Physics, CERN, 1211 Geneva 23, Switzerland}

\affiliation{Department of Physics, Faculty of Sciences, Ankara University, 06100,
Tandogan, Ankara, Turkey}

\author{G. Ünel}

\email{gokhan.unel@cern.ch}

\affiliation{Department of Physics, CERN, 1211 Geneva 23, Switzerland}

\affiliation{Department of Physics, University of California at Irvine, USA}
\begin{abstract}
Considering the present limits on the masses of fourth family quarks
from the Tevatron experiments, the fourth family quarks are expected
to have mass larger than the top quark. Due to their expected large
mass they could have different dynamics than the quarks of three families
of the Standard Model. The resonant production of the fourth family
up type quark $t'$ has been studied via anomalous production subprocess
$gq_{i}\to t'$ (where $q_{i}=u,c$) at the LHC with the center of
mass energy 10 TeV and 14 TeV. The signatures of such process are
discussed within the SM decay modes. The sensitivity to anomalous
coupling $\kappa/\Lambda=0.1$ TeV$^{-1}$ can be reached at $\sqrt{s}=10$
TeV and $L_{int}=100$ pb$^{-1}$. 
\end{abstract}


\maketitle

\section{introduction}

The number of fermion families in nature, the pattern of fermion masses
and the mixing angles in the quark/lepton sectors are two of the unanswered
questions in the Standard Model (SM). The repetition of quark and
lepton families remain a mystery as part of the flavor problem. On
theoretical grounds, the asymptotic freedom in the quantum chromodynamics
imposes an indirect bound for the number of the quark flavors which
should be less than eighteen. The electroweak precision measurements
done by the Large Electron Positron (LEP) experiments imply that the
number of light neutrinos ($m_{\nu}<45$ GeV) is equal to three \cite{PDG}.
The most recent analyses indicate that an additional family of heavy
fermions is not inconsistent with the precision electroweak data at
the available energies \cite{holdom,hung,kribs,okun,murdock,he01,ozcan09}.
Indeed, the presence of three or four fermion families are equally
consistent with the electroweak precision data, moreover the four
families scenario is favored if the Higgs boson heavier than $200$
GeV \cite{kribs}. The fourth family may play an important role in
our understanding of the flavor structure of the SM. The flavor democracy
is one of the main motivations for the existence of the fourth family
fermions \cite{4thfam}. Another motivation for the fourth family
comes from the charge-spin unification \cite{mankoc}. Additional
fermions can also be accommodated in many models beyond the SM \cite{Nath08}.\emph{
}A recent review of the fourth SM family including the theoretical
and experimental aspects can be found in \cite{Mangano09}.

A lower limit on the mass of the fourth family quark $Q'$ is $m_{Q'}>256$
GeV from Tevatron experiments \cite{R-CDF-t'}, whereas the upper
limit from partial wave unitarity is about 1 TeV. The recent results
from the Collider Detector at Fermilab (CDF) experiment exclude the
$t'$ mass below $311$ GeV at $95\%$ CL using the data of 2.8 fb$^{-1}$\cite{CDF-public}. 

The tree level flavor changing processes occur only via the charged
current interactions in the SM. The first two rows of the Cabibbo-Kobayashi-Maskawa
(CKM) matrix \cite{CKM} are in good agreement with the unitary condition.
The data for the number of $b-$jets in the top quark pair production
at Tevatron constrain the ratio $R=|V_{tb}|^{2}/\sum_{q}|V_{tq}|^{2}$,
which is closely related to $V_{tb}$ if the CKM is unitary. The direct
constraints on $|V_{tb}|$ come from the single production of top
quarks at the Tevatron. From the average cross section $\sigma=3.7\pm0.8$
pb the lower limit $|V_{tb}|>0.74$ at $95\%$ CL is given by the
CDF and D$0$\cite{D0CDF}. A measurement of the single top production
cross section smaller than the SM prediction would imply $V_{tb}<1$
or the evidence of extra families of quarks mixed with the third generation.
A flavor extension of the SM, with a fourth generation of quarks,
leads to an extended CKM matrix which could have $V_{tb}$ smaller
than $1$. In the extended model, the strongest constraint to $V_{tb}$
comes from the ratio $R_{b}=\Gamma(Z\to b\overline{b})/\Gamma(Z\to\mbox{hadrons})$,
with $R>0.9$ for $m_{t'}\geq1.5m_{t}$. For an extra up type quark
$t'$ and another extra down type quark $b'$, the $4\times4$ matrix
is unitary, for which any $3\times3$ submatrix becomes non-unitary
as long as these new quarks mix with quarks of three families. Hence,
the new flavor changing neutral currents (FCNC) could appear without
violating the existing bounds from current experimental measurements
\cite{Herrera08,bobrowski09}.

The existence of a fourth generation of quarks would have interesting
implications. Taking into account the current bounds on the mass of
the fourth family quarks \cite{PDG}, the anomalous interactions can
emerge in the fourth family case. Furthermore, extra families will
yield an essential enhancement in the Higgs boson production at the
LHC \cite{Arik02}. Single production \cite{Ciftci07}, \cite{OC08}
mechanism of fourth family quarks will be suppressed by the elements
(fourth row and/or fourth column) of the 4$\times$4 CKM matrix. The
fourth family quark pairs can already be produced at the Large Hadron
Collider (LHC) at an initial center of mass energy of $\sqrt{s}=10$
TeV and an initial luminosity of $L=10^{31}$ cm$^{-2}$s$^{-1}$.
At the nominal center of mass energy $\sqrt{s}=14$ TeV the initial
luminosity will be $10^{33}$cm$^{-2}$s$^{-1}$ which will later
increase to $10^{34}$cm$^{-2}$s$^{-1}$ corresponding to 10 and
100 fb$^{-1}$ per year, respectively. 

In this work, we present an analysis of the anomalous resonant production
of $t'$ quarks at the LHC. Here, we assume the case $t'$ decays
through SM dominated channel (via charged currents) in which the magnitude
of $V_{t'q}$ is important, leading to a final state $W^{\pm}b_{jet}$
for $t'$ anomalous production. A fast simulation is performed for
the detector effects on the signal and background. Any observations
of the invariant mass peak in the interval 300-800 GeV with the fi{}nal
state containing $W^{\pm}b_{jet}$ can be interpreted as the signal
for $t'$ anomalous resonant production.

\section{Fourth Fam\i{}ly Quark Interactions}

Fourth family quarks can couple to charged weak currents by the exchange
of $W^{\pm}$ boson, neutral weak currents by $Z^{0}$ boson exchange,
electromagnetic currents by photon exchange and strong colour currents
by the gluons. We include the fourth family quarks in the enlarged
framework (primed) of the SM. The interaction lagrangian is given
by\begin{eqnarray}
L' & =-g_{e} & \sum_{Q_{i}'=b',t'}Q_{ei}\overline{Q_{i}}'\gamma^{\mu}Q_{i}'A_{\mu}-g_{s}\sum_{Q_{i}'=b',t'}\overline{Q_{i}}'T^{a}\gamma^{\mu}Q_{i}'G_{\mu}^{a}\nonumber \\
 &  & -\frac{g_{e}}{2\cos\theta_{w}\sin\theta_{w}}\sum_{Q_{i}'=b',t'}\overline{Q}_{i}'\gamma^{\mu}(g_{V}^{i}-g_{A}^{i}\gamma^{5})Q_{i}'Z_{\mu}^{0}\nonumber \\
 &  & -\frac{g_{e}}{2\sqrt{2}\sin\theta_{w}}\sum_{Q'_{i\neq j}=b',t'}V_{ij}\overline{Q}_{i}'\gamma^{\mu}(1-\gamma^{5})q_{j}W_{\mu}^{\pm}+h.c.\label{eq:1}\end{eqnarray}
where $g_{e}$ is the electromagnetic coupling constant, $g_{s}$
is the strong coupling constant. The vector fields $A_{\mu}$, $G_{\mu}$,
$Z_{\mu}$ and $W_{\mu}^{\pm}$ denote photon, gluon, $Z^{0}-$boson
and $W^{\pm}-$boson, respectively. $Q_{ei}$ is the electric charge
of fourth family quarks, $T^{a}$ are the Gell-Mann matrices. The
$g_{V}$ and $g_{A}$ are the couplings for vector and axial-vector
neutral currents. Finally, the CKM matrix elements $V_{ij}$ are expressed
as: $V=V^{U}V^{D\dagger}.$ The corresponding $4\times4$ CKM matrix
is given by 

\begin{equation}
V=\left(\begin{array}{cccc}
\mathbf{V_{ud}} & \mathbf{V_{us}} & \mathbf{V_{ub}} & V_{ub'}\\
\mathbf{V_{cd}} & \mathbf{V_{cs}} & \mathbf{V_{cb}} & V_{cb'}\\
\mathbf{V_{td}} & \mathbf{V_{ts}} & \mathbf{V_{tb}} & V_{tb'}\\
V_{t'd} & V_{t's} & V_{t'b} & V_{t'b'}\end{array}\right)\label{eq:2}\end{equation}
The magnitude of the $3\times3$ CKM matrix elements are determined
from the low energy and high energy experiments: these are $|V_{ud}|=0.97418\pm0.00027$,
$|V_{us}|=0.2255\pm0.0019$, $|V_{cd}|=0.230\pm0.011$, $|V_{cs}|=1.04\pm0.06$,
$|V_{cb}|=0.0412\pm0.0011$, $|V_{ub}|=0.00393\pm0.00036$, $|V_{td}|=0.0081\pm0.0006$,
$|V_{ts}|=0.0387\pm0.0023$ (assuming $|V_{tb}|$ equal to unity)
and a lower limit from the single top production $|V_{tb}|>0.74$
at $95\%$ CL. \cite{PDG}. In the fourth family case, the first three
rows of this matrix are calculated as $|V_{ub'}|^{2}=0.0008$, $|V_{cb'}|^{2}=0.0295$
and $|V_{tb'}|^{2}=0.4054$. For the first three columns one calculates
$|V_{t'd}|^{2}=0.001$, $|V_{t's}|^{2}=0.0315$ and $|V_{t'b}|^{2}=0.4053$.
We see that there is a loose constraint for the mixing between third
and fourth family quarks. In this case, these bounds can be relaxed
to an uncertainty level. If there is a mass degeneracy between $t'$
and $b'$ quarks, the two body decays occur most probably into the
third family quarks. Inspiring from the Wolfenstein parametrization
of the $3\times3$ CKM matrix, we could simply consider the fourth
row and fourth column of the $4\times4$ CKM as $|V_{q_{i}b'}|\simeq|V_{t'q_{j}}|=A_{ij}\lambda^{4-n}$
where $A_{ij}$ can be optimized for the quark flavors $q_{i}$ and
$q_{j}$; $n$ is the family number and $\lambda$ is a constant.

We consider the decay width of $t'$ quark through $t'\to W^{+}q$
including the final state quark mass, we find 

\begin{equation}
\Gamma(Q'\to Wq)=\frac{1}{16}\frac{\alpha_{e}|V_{Q'q}|^{2}m_{Q'}^{3}}{\sin^{2}\theta_{W}m_{W}^{2}}\lambda_{W}\sqrt{\lambda_{r}}\label{eq:4}\end{equation}
where 

\begin{equation}
\lambda_{r}=1+m_{W}^{4}/m_{Q'}^{4}+m_{q}^{4}/m_{Q'}^{4}-2m_{W}^{2}/m_{Q'}^{2}-2m_{q}^{2}/m_{Q'}^{2}-2m_{W}^{2}m_{q}^{2}/m_{Q'}^{4}\label{eq:5}\end{equation}

\begin{equation}
\lambda_{W}=1+m_{W}^{2}/M_{Q'}^{2}-2m_{q}^{2}/m_{Q'}^{2}+m_{q}^{4}/m_{Q'}^{4}+m_{q}^{2}m_{W}^{2}/m_{Q'}^{4}-2m_{W}^{4}/m_{Q'}^{4}\label{eq:6}\end{equation}

To calculate the decay width numerically, we assume three parametrizations
PI, PII and PIII for the fourth family mixing matrix elements. For
the PI parametrization we assume the constant values $|V_{Q'q}|=|V_{qQ'}|=0.01$,
PII contains a dynamical parametrization $|V_{q_{i}Q'}|=|V_{Q'q_{j}}|\approx\lambda^{4-n}$
with a preferred value of $\lambda=0.1$, and PIII has the parameters
$|V_{t'd}|=0.063$, $|V_{t's}|=0.46$, $|V_{t'b}|=0.47$, $|V_{ub'}|=0.044$,
$|V_{cb'}|=0.46$, $|V_{tb'}|=0.47$ \cite{pair_search}.

The flavor changing neutral current interactions are known to be absent
at tree level in the SM. However, the fourth family quarks, being
heavier than the top quark, could have different dynamics than other
quarks and they can couple to the FCNC currents leading to an enhancement
in the resonance processes at the LHC. Moreover, the arguments for
the anomalous interactions of the top quark given in \cite{fritzsch},
are more valid for $t'$ and $b'$ quarks. The effective Lagrangian
for the anomalous interactions among the fourth family quarks $t'$
and $b'$, ordinary quarks $q$, and the neutral gauge bosons $V=\gamma,Z,g$
can be written explicitly:

\begin{eqnarray}
L'_{a} & = & \sum_{q_{i}=u,c,t}\frac{\kappa_{\gamma}^{q_{i}}}{\Lambda}Q_{q_{i}}g_{e}\overline{t}'\sigma_{\mu\nu}q_{i}F^{\mu\nu}+\sum_{q_{i}=u,c,t}\frac{\kappa_{z}^{q_{i}}}{2\Lambda}g_{z}\overline{t}'\sigma_{\mu\nu}q_{i}Z^{\mu\nu}\nonumber \\
 &  & +\sum_{q_{i}=u,c,t}\frac{\kappa_{g}^{q_{i}}}{2\Lambda}g_{s}\overline{t}'\sigma_{\mu\nu}\lambda_{a}q_{i}G_{a}^{\mu\nu}+h.c.\nonumber \\
 &  & +\sum_{q_{i}=d,s,b}\frac{\kappa_{\gamma}^{q_{i}}}{\Lambda}Q_{q_{i}}g_{e}\overline{b}'\sigma_{\mu\nu}q_{i}F^{\mu\nu}+\sum_{q_{i}=d,s,b}\frac{\kappa_{z}^{q_{i}}}{2\Lambda}g_{z}\overline{b}'\sigma_{\mu\nu}q_{i}Z^{\mu\nu}\nonumber \\
 &  & +\sum_{q_{i}=d,s,b}\frac{\kappa_{g}^{q_{i}}}{2\Lambda}g_{s}\overline{b}'\sigma_{\mu\nu}\lambda_{a}q_{i}G_{a}^{\mu\nu}+h.c.\label{eq:3}\end{eqnarray}
where $F^{\mu\nu}$, $Z^{\mu\nu}$ and $G^{\mu\nu}$ are the fi{}eld
strength tensors of the gauge bosons; $\sigma_{\mu\nu}=i(\gamma_{\mu}\gamma_{\nu}-\gamma_{\nu}\gamma_{\mu})/2$;
$\lambda_{a}$ are the Gell-Mann matrices; $Q_{q}$ is the electric
charge of the quark ($q$); $g_{e}$, $g_{Z}$ and $g_{s}$ are the
electromagnetic, neutral weak and the strong coupling constants, respectively.
$g_{Z}=g_{e}/\cos\theta_{w}\sin\theta_{w}$, where $\theta_{w}$ is
the weak angle. $\kappa_{\gamma}$ is the anomalous coupling with
photon; $\kappa_{z}$ is for the $Z$ boson, and $\kappa_{g}$ with
gluon. $\Lambda$ is the cut-off scale for the new interactions. 

For the decays $Q'\to Vq$ where $V\equiv\gamma,Z,g$, we use the
effective Lagrangian to calculate the anomalous decay widths 

\begin{equation}
\Gamma(Q'\to gq)=\frac{2}{3}\left(\frac{\kappa_{g}^{q}}{\Lambda}\right)^{2}\alpha_{s}m_{Q'}^{3}\lambda_{0}\label{eq:7}\end{equation}

\begin{equation}
\Gamma(Q'\to\gamma q)=\frac{1}{2}\left(\frac{\kappa_{\gamma}^{q}}{\Lambda}\right)^{2}\alpha_{e}Q_{q}^{2}m_{Q'}^{3}\lambda_{0}\label{eq:8}\end{equation}

\begin{equation}
\Gamma(Q'\to Zq)=\frac{1}{16}\left(\frac{\kappa_{Z}^{q}}{\Lambda}\right)^{2}\frac{\alpha_{e}m_{Q'}^{3}}{\sin^{2}\theta_{W}\cos^{2}\theta_{W}}\lambda_{Z}\sqrt{\lambda_{r}}\label{eq:9}\end{equation}
with 

\begin{equation}
\lambda_{0}=1-3m_{q}^{2}/m_{Q'}^{2}+3m_{q}^{4}/m_{Q'}^{4}-m_{q}^{6}/m_{Q'}^{6}\label{eq:10}\end{equation}

\begin{equation}
\lambda_{Z}=2-m_{Z}^{2}/m_{Q'}^{2}-4m_{q}^{2}/m_{Q'}^{2}+2m_{q}^{4}/m_{Q'}^{4}-6m_{q}m_{Z}^{2}/m_{Q'}^{3}-m_{Z}^{2}m_{t}^{2}/m_{Q'}^{4}-m_{Z}^{4}/m_{Q'}^{4}\label{eq:11}\end{equation}

The anomalous decay widths in different channels are proportional
to $\Lambda^{-2}$, and they become to contribute more $\kappa/\Lambda>0.1$
TeV$^{-1}$. 

\begin{table}
\caption{Decay widths and branching ratios ($\%$) of the fourth family quarks
in case both chiral and anomalous interactions, where $q_{i}$ denotes
the quarks belong to $i^{th}$ family. $Vq_{1,2}$ denotes $Vq_{1}$
and $Vq_{2}$. The parametrization for the $4\times4$ CKM elements
is taken as PI, PII, PIII given in the text. In the last row the total
decay widths of fourth family $t'$ quarks are given for three mass
values and $\kappa/\Lambda=1$ TeV$^{-1}$. The numbers in the paranthesis
denote the values for $\kappa/\Lambda=0.1$ TeV$^{-1}$.\label{tab:1}}

\begin{tabular}{|c|c|c|c|c|c|c|c|c|c|}
\hline 
 & \multicolumn{3}{c|}{PI} & \multicolumn{3}{c|}{PII} & \multicolumn{3}{c|}{PIII}\tabularnewline
\hline
$m_{t'}$(GeV) & 300 & 500 & 700 & 300 & 500 & 700 & 300 & 500 & 700\tabularnewline
\hline 
$Wq_{1}$ & 0.017(1.6) & 0.014(1.4) & 0.014(1.3) & 0.0002(0.0062) & 0.0001(0.0059) & 0.0001(0.0058) & 0.39(0.9) & 0.35(0.9) & 0.34(0.89)\tabularnewline
\hline 
$Wq_{2}$ & 0.017(1.6) & 0.014(1.4) & 0.014(1.3) & 0.017(0.62) & 0.014(0.59) & 0.014(0.58) & 21.0(48) & 19.0(48) & 18.0(48)\tabularnewline
\hline 
$Wq_{3}$ & 0.017(1.6) & 0.014(1.4) & 0.014(1.3) & 1.7(62) & 1.4(59) & 1.4(58) & 21.0(50) & 20.0(50) & 19.0(50)\tabularnewline
\hline 
$Zq_{1,2}$ & 2.5(2.3) & 2.3(2.2) & 2.2(2.1) & 2.4(0.91) & 2.3(0.93) & 2.2(0.93) & 1.4(0.033) & 1.4(0.036) & 1.4(0.037)\tabularnewline
\hline
$Zq_{3}$ & 0.27(0.26) & 1.4(1.4) & 1.8(1.7) & 0.27(0.1) & 1.4(0.59) & 1.8(0.75) & 0.16(0.0036) & 0.89(0.023) & 1.1(0.03)\tabularnewline
\hline 
$\gamma q_{1,2}$ & 0.9(0.86) & 0.76(0.73) & 0.72(0.69) & 0.89(0.33) & 0.75(0.31) & 0.71(0.3) & 0.52(0.012) & 0.47(0.012) & 0.45(0.012)\tabularnewline
\hline 
$\gamma q_{3}$ & 0.26(0.25) & 0.52(0.5) & 0.6(0.57) & 0.26(0.097) & 0.51(0.21) & 0.59(0.25) & 0.52(0.0035) & 0.32(0.008) & 0.37(0.0098)\tabularnewline
\hline 
$gq_{1,2}$ & 40(39) & 34(33) & 32(31) & 40(15) & 34(14) & 32(14) & 23.0(0.54) & 21.0(0.53) & 20.0(0.53)\tabularnewline
\hline
$gq_{3}$ & 12(11) & 23(22) & 27(26) & 12(4.4) & 23(9.4) & 26(11) & 6.8(0.16) & 14.0(0.36) & 20.0(0.44)\tabularnewline
\hline
$\Gamma_{tot}$(GeV) & 5.21(0.055) & 28.47(0.297) & 82.58(0.859) & 5.298(0.141) & 28.871(0.701) & 83.71(1.97) & 9.05(3.89) & 46.46(18.29) & 132.04(50.30)\tabularnewline
\hline
\end{tabular}
\end{table}

\begin{figure}
\includegraphics[scale=0.5]{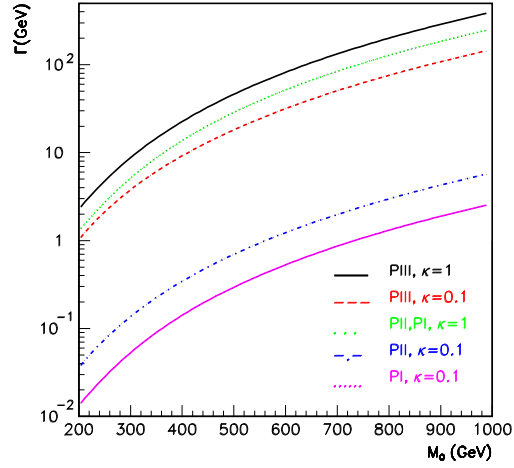}

\caption{Decay width of $t'$ quark depending on its mass for different $V_{t'q}$
parametrizations and $\kappa/\Lambda$ values. \label{fig:1}}

\end{figure}

The total decay width of $t'$ quark is shown in Fig.\ref{fig:1}.
In Table \ref{tab:1}, we give the numerical values of the total decay
width and branching ratios for the parametrizations PI-PIII. As it
can be seen from these tables, for the mass range of $m_{t'}$ relevant
to LHC experiments, the fraction of the anomalous decay modes are
$99.9\%$($95-96\%$), $98\%$($37-41\%$) and $58-62\%$($99\%$)
at $\kappa/\Lambda=1(0.1)$ TeV$^{-1}$ for the parametrizations PI,
PII and PIII, respectively. However, SM decay modes of $t'$ become
dominant at PII and PIII parametrizations when $\kappa/\Lambda=0.1$
TeV$^{-1}$. For the parametrization PIII, the SM decay mode and anomalous
decay mode become comparable for $\kappa/\Lambda=1$ TeV$^{-1}$.

If the CKM unitarity is strictly applied and the $Q'$ mixing with
light quarks is strongly constrained, which corresponds to $|V_{t'q_{i}}|\simeq0$
for the light quarks $q_{i}$, there still remains room for the SM
decays $t'\to W^{+}b$ and the possible anomalous decays $t'\to Vq_{i}$.
In this case, $t'$ can have significant FCNC couplings. When the
anomalous interactions much dominate over the SM decays, still allowing
only the non-vanishing element $|V_{t'b'}|\simeq1$ (but assuming
$t'$ and $b'$ are mass degenerate) the decay widths and branching
ratios of the fourth family quarks with the anomalous interactions
are shown in Table \ref{tab:2}. Taking the anomalous coupling $\kappa/\Lambda=0.1$
TeV$^{-1}$, we calculate the $t'$ anomalous decay width $\Gamma=8.79\times10^{-2}$
GeV, $\Gamma=3.86\times10^{-1}$ GeV and $\Gamma=1.02$ GeV for $m_{t'}=350$,
$550$ and $750$ GeV, respectively. 

\begin{table}
\caption{Decay widths and branching ratios ($\%$) of the fourth family quarks
with only anomalous interactions, where $q_{i}$ denotes the quarks
belong to $i^{th}$ family. $Vq_{1,2}$ denotes $Vq_{1}$ and $Vq_{2}$.
In the last row the total decay widths of fourth family $t'$ and
$b'$ quarks are given for three mass values and $\kappa/\Lambda$=1
TeV$^{-1}$. \label{tab:2}}

\begin{tabular}{|c|c|c|c|}
\hline 
 & \multicolumn{3}{c|}{$t'$}\tabularnewline
\hline
Mass (GeV) & 300 & 500 & 700\tabularnewline
\hline 
$Zq_{1,2}$ & 2.5 & 2.3 & 2.2\tabularnewline
\hline
$Zq_{3}$ & 0.27 & 1.4 & 1.8\tabularnewline
\hline 
$\gamma q_{1,2}$ & 0.9 & 0.76 & 0.72\tabularnewline
\hline 
$\gamma q_{3}$ & 0.26 & 0.52 & 0.6\tabularnewline
\hline 
$gq_{1,2}$ & 40.0 & 34.0 & 32.0\tabularnewline
\hline
$gq_{3}$ & 12.0 & 23.0 & 27.0\tabularnewline
\hline
$\Gamma_{tot}$(GeV) & 5.21 & 28.4 & 82.56\tabularnewline
\hline
\end{tabular}
\end{table}

\section{Anomalous Resonant Production of $t'$ Quarks}

\begin{figure}
\begin{centering}
\includegraphics{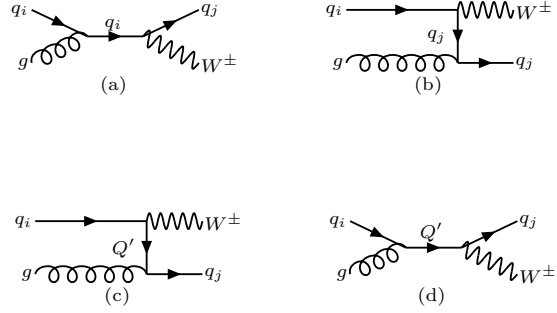}
\par\end{centering}

\caption{Diagrams for the $gq_{i}\to W^{+}q_{j}$ subprocess including $Q'qg$
(where $Q'\equiv t'$) anomalous vertices; $q_{i}$ can be any of
the quarks inside the proton, while $q_{j}$ can be any of the quark
flavour depending on the charged current interaction. For the $W^{-}\bar{q}_{j}$
final states we may change the direction of the current lines and
replace the incoming anti-quarks. \label{fig:2}}

\end{figure}

In order to study the resonant production of fourth family quarks,
we have implemented the anomalous interaction vertices with the new
particles $t'$ into CompHEP package \cite{comphep}. In all numerical
calculations, the parton distribution functions (PDF) $f_{q}(x,Q^{2})$
and $f_{g}(y,Q^{2})$ are set to the CTEQ6M parametrization \cite{R-cteq}
and the factorization scale $Q^{2}=m_{Q'}^{2}$ is used. The total
cross section for the process $pp\to W^{+}bX$ is given by

\begin{equation}
\sigma=\int_{\tau_{min}}^{1}d\tau\int_{\tau}^{1}\frac{dx}{x}f_{q}(x,Q^{2})f_{g}(\frac{\tau}{x},Q^{2})\hat{\sigma}(\tau s)\label{eq:12}\end{equation}
where $\hat{s}=\tau s$ and $\tau_{min}=(m_{V}+m_{q})^{2}/s$; $\hat{\sigma}$
is the partonic cross section for a given process. We consider the
production $qg\rightarrow t'$ and the decay $t\rightarrow W^{+}b_{j}$.
In the generic notation the contributing Feynman diagrams are shown
in Fig.\ref{fig:2}.

\begin{figure}
\includegraphics[scale=0.6]{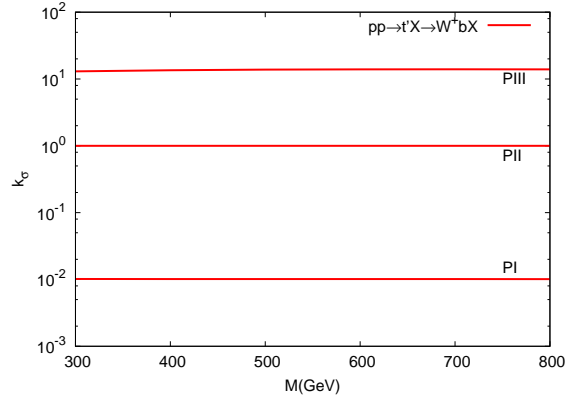}

\caption{The ratio of the cross sections for different parametrizations normalized
to PII, where $\kappa/\Lambda=1$ TeV$^{-1}$.\label{fig:3}}

\end{figure}

\begin{figure}
\begin{centering}
\includegraphics[scale=0.4]{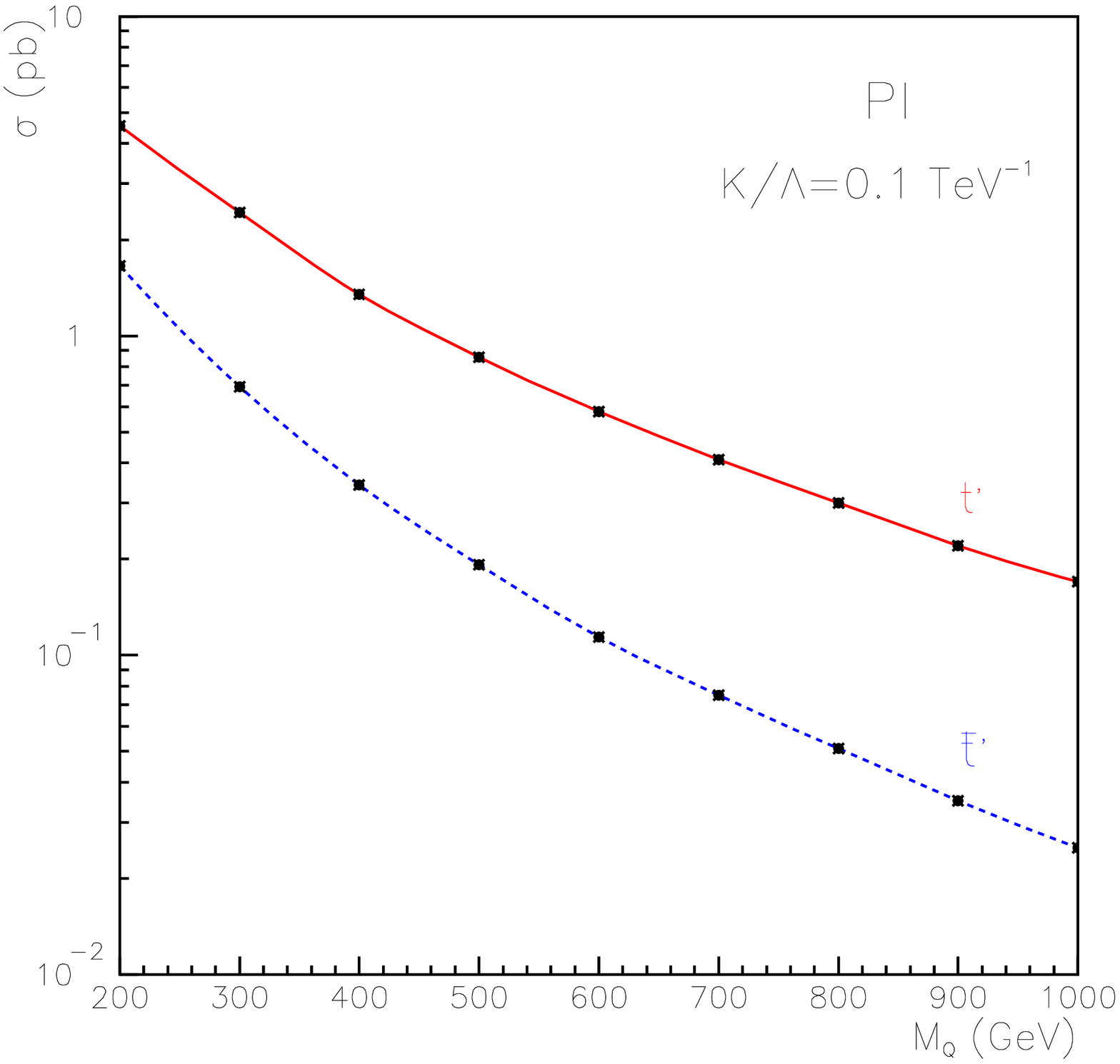}\includegraphics[scale=0.4]{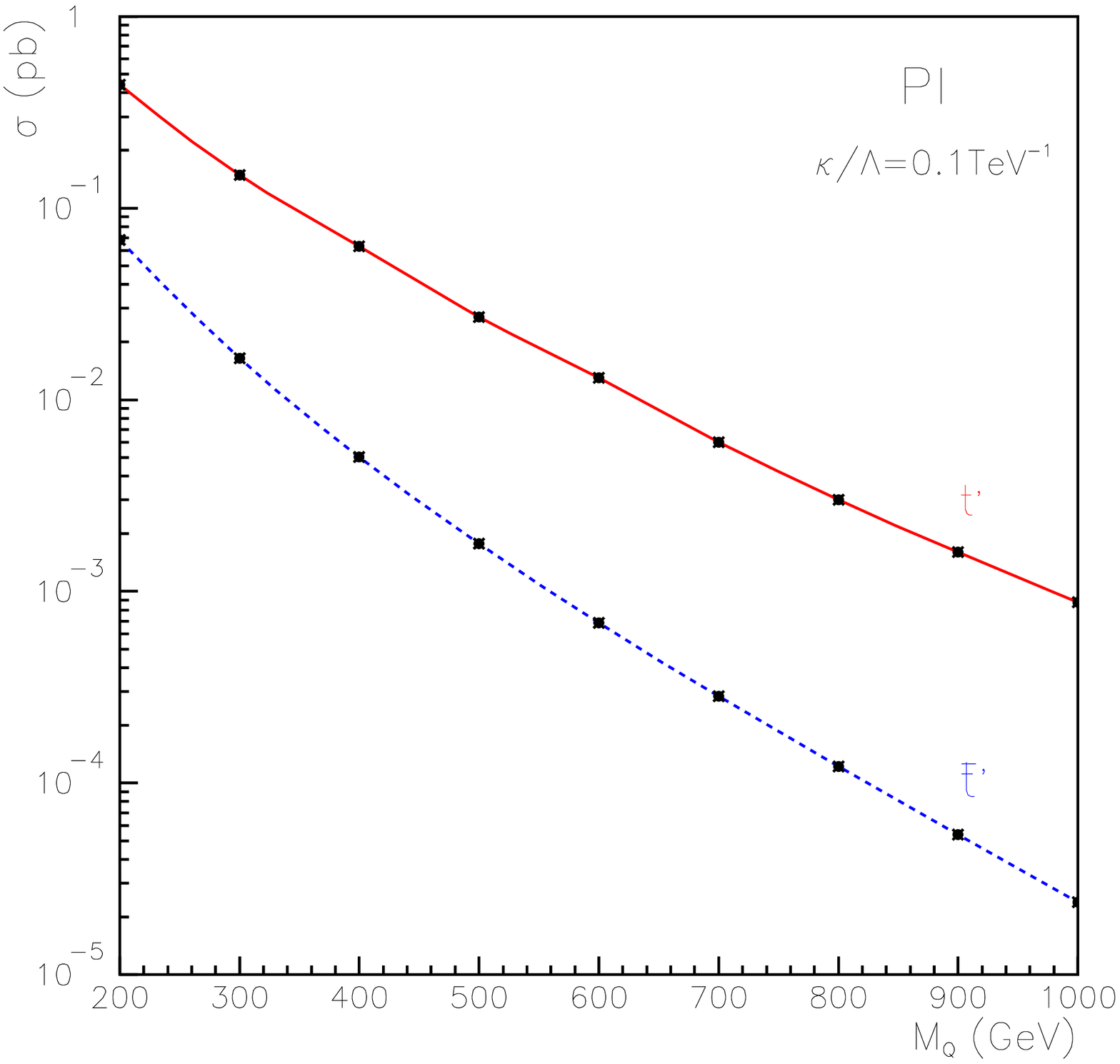}
\par\end{centering}

\caption{The cross section for the anomalous production of $t'$ quarks for
PI parametrization and $\kappa/\Lambda=0.1$ TeV$^{-1}$. The lines
denotes $t'(\bar{t'})$ followed by the decay $W^{+}b$($W^{-}\bar{b}$)
at the center of mass energy 14 TeV (left) and 10 TeV (right).\label{fig:4}}

\end{figure}

\begin{figure}
\begin{centering}
\includegraphics[scale=0.4]{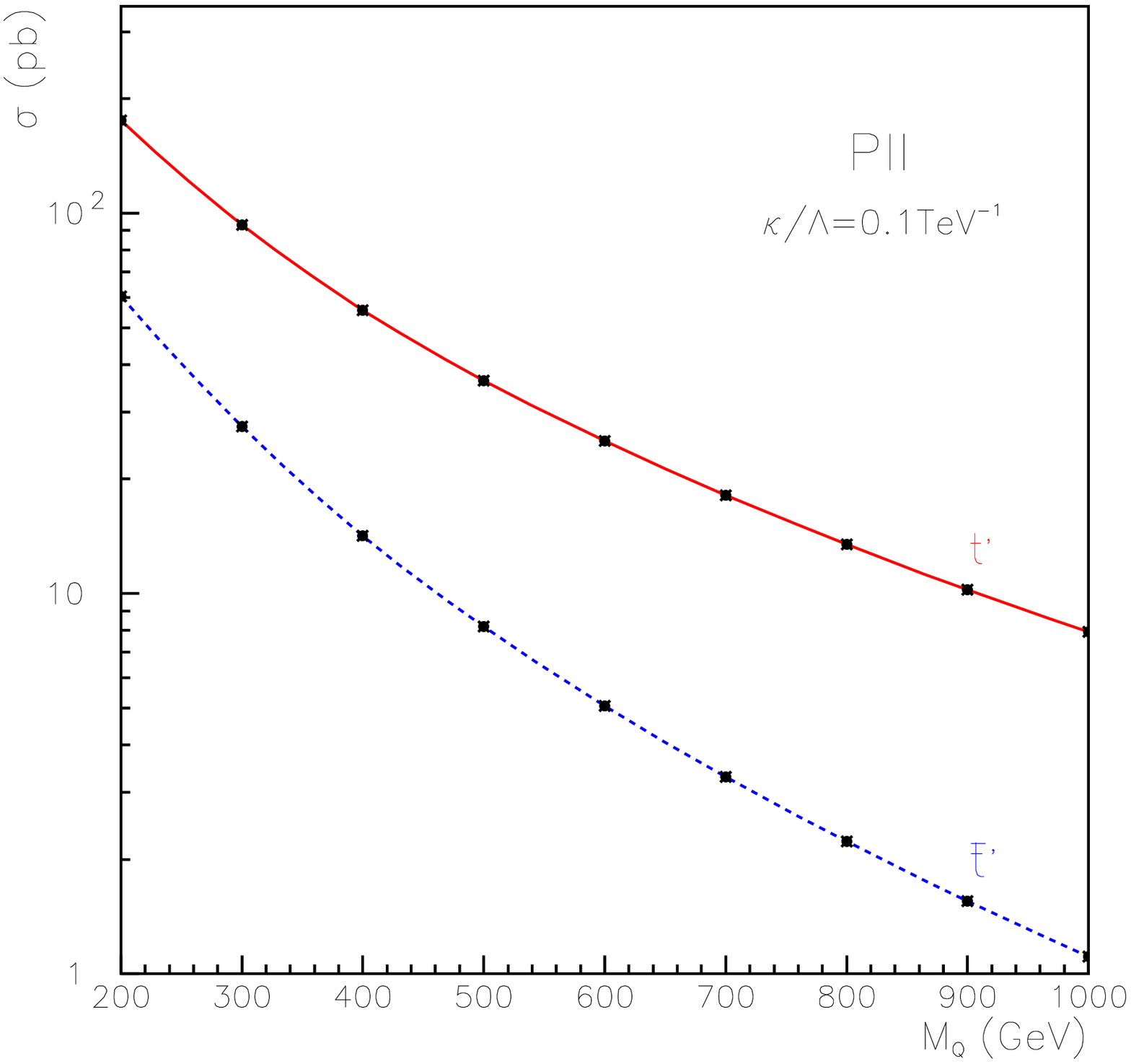}\includegraphics[scale=0.4]{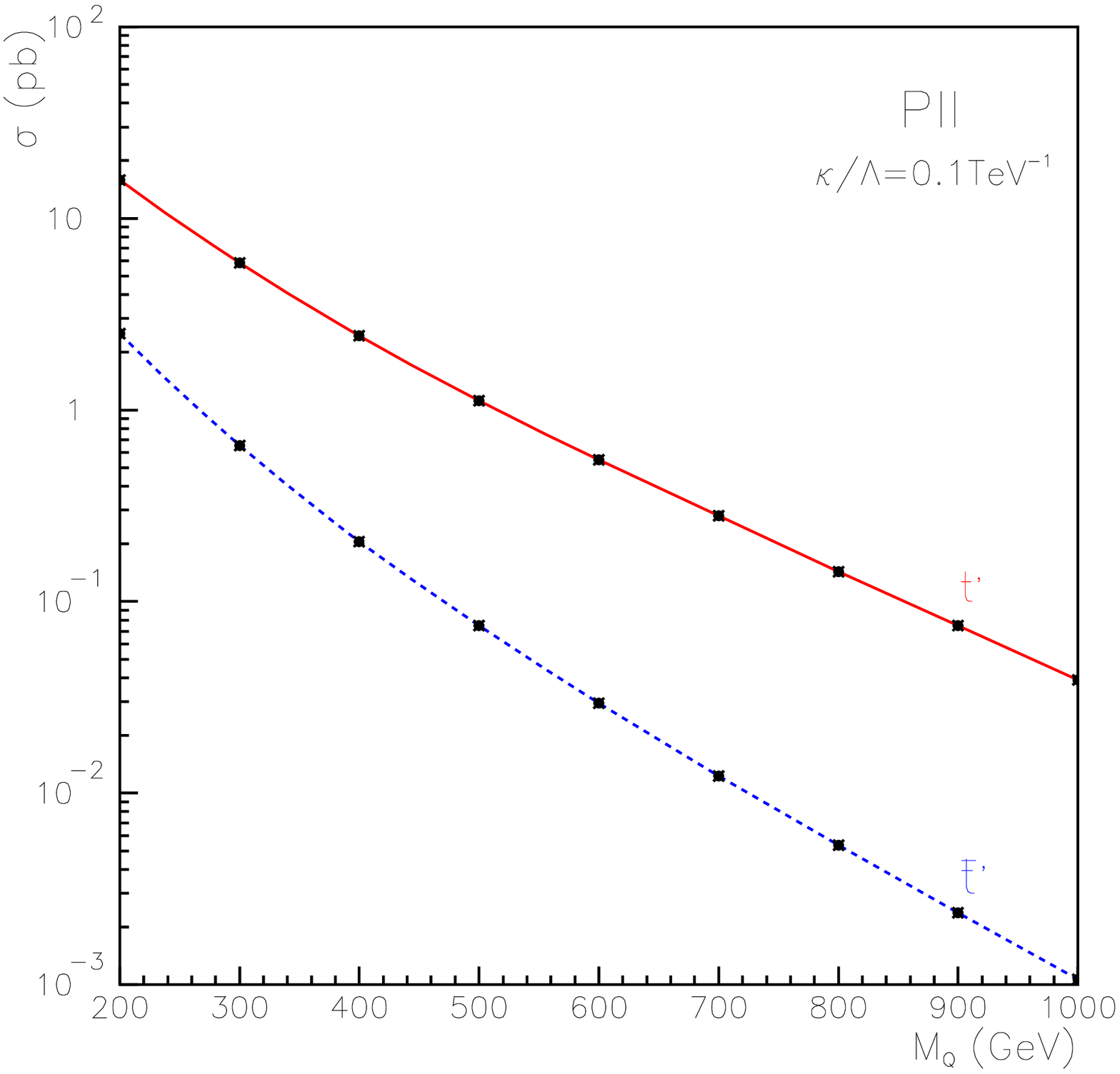}
\par\end{centering}

\caption{The cross section for the anomalous production of $t'$ quarks for
PII parametrization and $\kappa/\Lambda=0.1$ TeV$^{-1}$. The lines
denote $t'(\bar{t'})$ production followed by the decay $W^{+}b$($W^{-}\bar{b}$)
at the center of mass energy 14 TeV (left) and 10 TeV (right). \label{fig:5}}

\end{figure}

\begin{figure}
\begin{centering}
\includegraphics[scale=0.4]{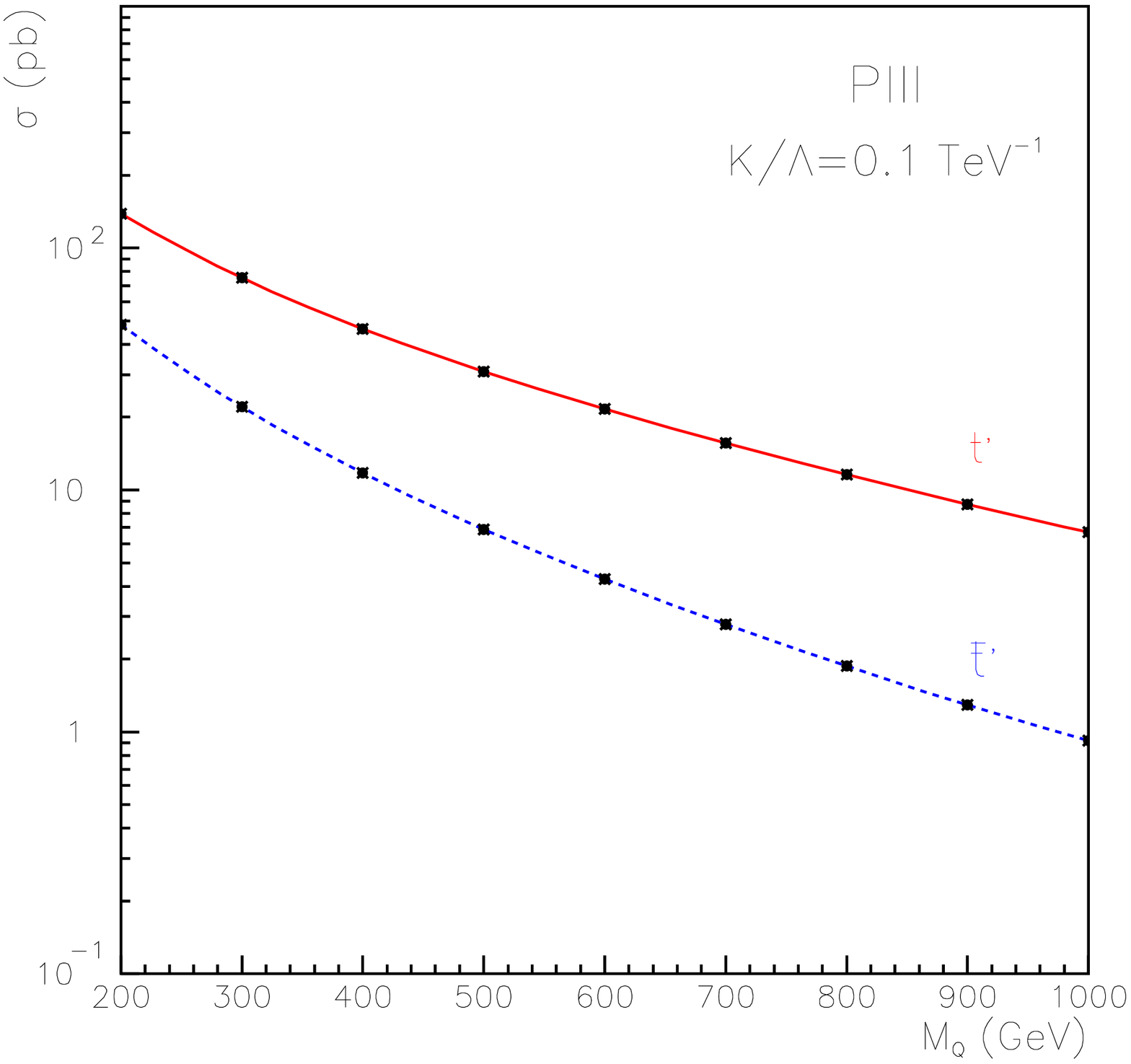}\includegraphics[scale=0.4]{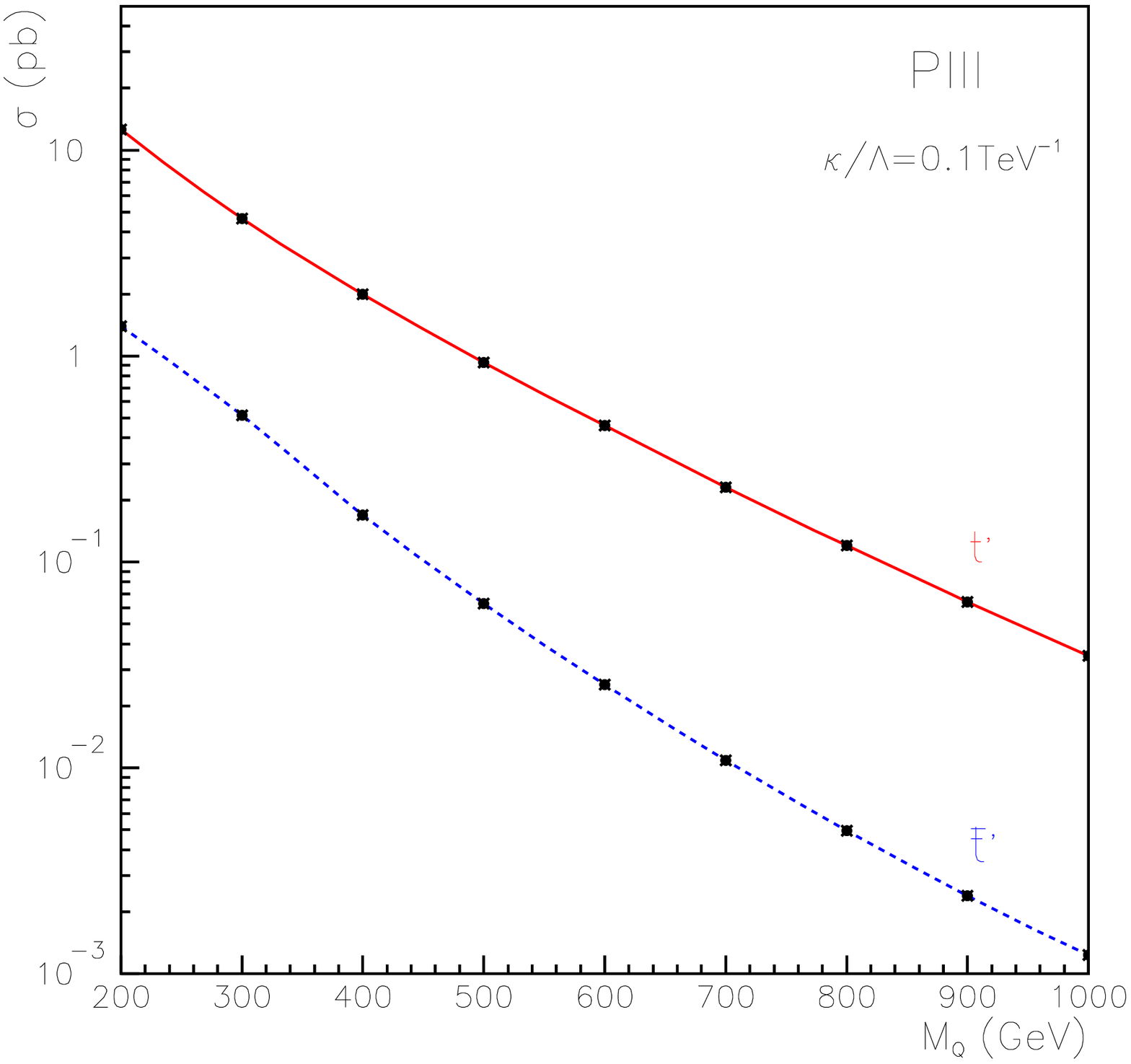}
\par\end{centering}

\caption{The cross section for the anomalous production of $t'$ quarks for
PIII parametrization and $\kappa/\Lambda=0.1$ TeV$^{-1}$. The lines
denote $t'(\bar{t'})$ production followed by the decay $W^{+}b$($W^{-}\bar{b}$)
at the center of mass energy 14 TeV (left) and 10 TeV (right). \label{fig:6}}

\end{figure}

The production cross sections as a function of fourth family quark
mass for the different parametrization are shown in Tables \ref{tab:3}-\ref{tab:5}.
The ratios of the cross sections for different parametrizations are
calculated as $k_{\sigma}\approx0.01$ for PI and $k_{\sigma}\approx15$
for PIII with the normalization to PII with $\kappa/\Lambda=1$ TeV$^{-1}$,
as shown in Fig. \ref{fig:3}. For the parametrization PIII we find
the $t'(\overline{t}')$ production cross sections $46.4(11.7)$ pb
for $\kappa/\Lambda=0.1$ TeV$^{-1}$ and $m_{t'}=400$ GeV at $\sqrt{s}=14$
TeV. The $\overline{t}'$ production cross section is lower than $t'$
production with a factor of 2-8 depending on the considered mass range.
The general behaviour of the cross sections depending on the mass
is presented in Figs.\ref{fig:4}-\ref{fig:6}.

\begin{table}
\caption{Resonance cross sections (pb) and decay widths of $t'$ quarks for
PI parametrization with $\kappa/\Lambda=1$ TeV$^{-1}$ and $\kappa/\Lambda=0.1$
TeV$^{-1}$ at $\sqrt{s}=14(10)$ TeV.\label{tab:3}}

\begin{tabular}{|c|c|c|c|c|c|c|}
\hline 
 & \multicolumn{3}{c|}{$\kappa/\Lambda=1$ TeV$^{-1}$} & \multicolumn{3}{c|}{$\kappa/\Lambda=0.1$ TeV$^{-1}$}\tabularnewline
\hline 
$m_{t'}$(GeV) & $\sigma(t')$ & $\sigma(\bar{t'})$ & $\Gamma_{t'}$(GeV) & $\sigma(t')$ & $\sigma(\bar{t'})$ & $\Gamma_{t'}$(GeV)\tabularnewline
\hline
\hline 
300 & 2.55(0.16) & 0.744(1.73x10$^{-3}$) & 5.21 & 2.53(0.149) & 0.693(1.65x10$^{-2}$) & 0.055\tabularnewline
\hline 
400 & 1.43(6.11x10$^{-2}$) & 0.360(5.16x10$^{-3}$) & 13.74 & 1.34(0.063) & 0.341(5.02x10$^{-3}$) & 0.143\tabularnewline
\hline 
500 & 0.903(2.68x10$^{-2}$) & 0.198(1.81x10$^{-3}$) & 28.46 & 0.856(0.027) & 0.192(1.77x10$^{-3}$) & 0.296\tabularnewline
\hline 
600 & 0.608(1.26x10$^{-2}$) & 0.119(7.07x10$^{-4}$) & 50.91 & 0.582(0.013) & 0.114(6.84x10$^{-4}$) & 0.530\tabularnewline
\hline 
700 & 0.429(6.25x10$^{-3}$) & 0.075(3.0x10$^{-4}$) & 82.59 & 0.415(0.006) & 0.075(2.84x10$^{-4}$) & 0.859\tabularnewline
\hline 
800 & 0.311(3.2x10$^{-3}$) & 0.049(1.38x10$^{-4}$) & 125.06 & 0.305(0.003) & 0.051(1.22x10$^{-4}$) & 1.30\tabularnewline
\hline 
900 & 0.232(1.69x10$^{-3}$) & 0.034(6.82x10$^{-5}$) & 179.82 & 0.217(1.7x10$^{-3}$) & 0.035(5.38x10$^{-5}$) & 1.86\tabularnewline
\hline 
1000 & 0.174(9.24x10$^{-4}$) & 0.023(3.66x10$^{-5}$) & 248.43 & 0.178(8.8x10$^{-4}$) & 0.025(2.39x10$^{-5}$) & 2.58\tabularnewline
\hline
\end{tabular}
\end{table}

\begin{table}
\caption{Resonance cross sections (pb) and decay widths of $t'$ quarks for
PII parametrization with $\kappa/\Lambda=1$ TeV$^{-1}$and $\kappa/\Lambda=0.1$
TeV$^{-1}$ at $\sqrt{s}=14(10)$ TeV.\label{tab:4}}

\begin{tabular}{|c|c|c|c|c|c|c|}
\hline 
 & \multicolumn{3}{c|}{$\kappa/\Lambda=1$ TeV$^{-1}$} & \multicolumn{3}{c|}{$\kappa/\Lambda=0.1$ TeV$^{-1}$}\tabularnewline
\hline 
$m_{t'}$(GeV) & $\sigma(t')$  & $\sigma(\overline{t}')$  & $\Gamma_{t'}$(GeV) & $\sigma(t')$  & $\sigma(\overline{t}')$  & $\Gamma_{t'}$(GeV)\tabularnewline
\hline
\hline 
300 & 250.92(15.48) & 73.30(1.71) & 5.29 & 93.24(5.87) & 27.51(0.65) & 0.14\tabularnewline
\hline 
400 & 141.24(6.02) & 35.53(0.52) & 13.94 & 55.62(2.44) & 14.18(0.21) & 0.35\tabularnewline
\hline 
500 & 88.89(2.64) & 19.61(0.18) & 28.87 & 36.26(1.12) & 8.18(7.48x10$^{-2}$) & 0.70\tabularnewline
\hline 
600 & 60.00(1.25) & 11.79(6.98x10$^{-2}$) & 51.61 & 25.16(0.55) & 5.06(2.95x10$^{-2}$) & 1.23\tabularnewline
\hline 
700 & 42.33(0.62) & 7.46(2.96x10$^{-2}$) & 83.71 & 18.14(0.28) & 3.29(1.23x10$^{-2}$) & 1.97\tabularnewline
\hline 
800 & 30.77(0.32) & 4.92(1.36x10$^{-2}$) & 126.72 & 13.45(0.14) & 2.23(5.35x10$^{-3}$) & 2.96\tabularnewline
\hline 
900 & 22.84(0.17) & 3.34(6.76x10$^{-3}$) & 182.18 & 10.22(0.075) & 1.55(2.37x10$^{-3}$) & 4.23\tabularnewline
\hline 
1000 & 17.23(9.1x10$^{-2}$) & 2.32(3.6x10$^{-3}$) & 251.67 & 7.92(0.039) & 1.11(1.07x10$^{-3}$) & 5.82\tabularnewline
\hline
\end{tabular}
\end{table}

\begin{table}
\caption{Resonance cross sections (pb) and decay widths of $t'$ quarks for
PIII parametrization with $\kappa/\Lambda=1$TeV$^{-1}$and $\kappa/\Lambda=0.1$
TeV$^{-1}$ at $\sqrt{s}=14(10)$ TeV. \label{tab:5}}

\begin{tabular}{|c|c|c|c|c|c|c|}
\hline 
 & \multicolumn{3}{c|}{$\kappa/\Lambda=1$ TeV$^{-1}$} & \multicolumn{3}{c|}{$\kappa/\Lambda=0.1$ TeV$^{-1}$}\tabularnewline
\hline
\hline 
$m_{t'}$(GeV) & $\sigma(t')$ & $\sigma(\bar{t'})$ & $\Gamma_{t'}$(GeV) & $\sigma(t')$ & $\sigma(\bar{t'})$ & $\Gamma_{t'}$(GeV)\tabularnewline
\hline
\hline 
300 & 3272.60(198.72) & 946.54(21.92) & 9.05 & 75.28(4.66) & 22.03(0.515) & 3.89\tabularnewline
\hline 
400 & 1912.50(79.96)  & 475.42(6.75) & 22.92 & 46.43(2.00) & 11.74(0.169) & 9.33\tabularnewline
\hline 
500 & 1228.70(35.57)  & 267.17(2.42) & 46.46 & 30.89(0.93) & 6.87(6.26x10$^{-2}$) & 18.28\tabularnewline
\hline 
600 & 835.63(16.85) & 161.31(0.96) & 82.03 & 21.60(0.46) & 4.28(2.53x10$^{-2}$) & 31.65\tabularnewline
\hline 
700 & 590.53(8.34) & 102.24(0.42) & 132.04 & 15.61(0.23) & 2.78(1.1x10$^{-2}$) & 50.30\tabularnewline
\hline 
800 & 428.19(4.27) & 67.04(0.20) & 198.88 & 11.56(0.12) & 1.87(4.95x10$^{-3}$) & 75.12\tabularnewline
\hline 
900 & 315.96(2.26) & 45.14(9.96x10$^{-2}$) & 284.95 & 8.73(0.064) & 1.29(2.39x10$^{-3}$) & 106.99\tabularnewline
\hline 
1000 & 236.33(1.25) & 30.94(5.53x10$^{-2}$) & 392.65 & 6.69(0.035) & 0.92(1.23x10$^{-3}$) & 146.79\tabularnewline
\hline
\end{tabular}
\end{table}

\begin{figure}
\includegraphics[scale=0.8]{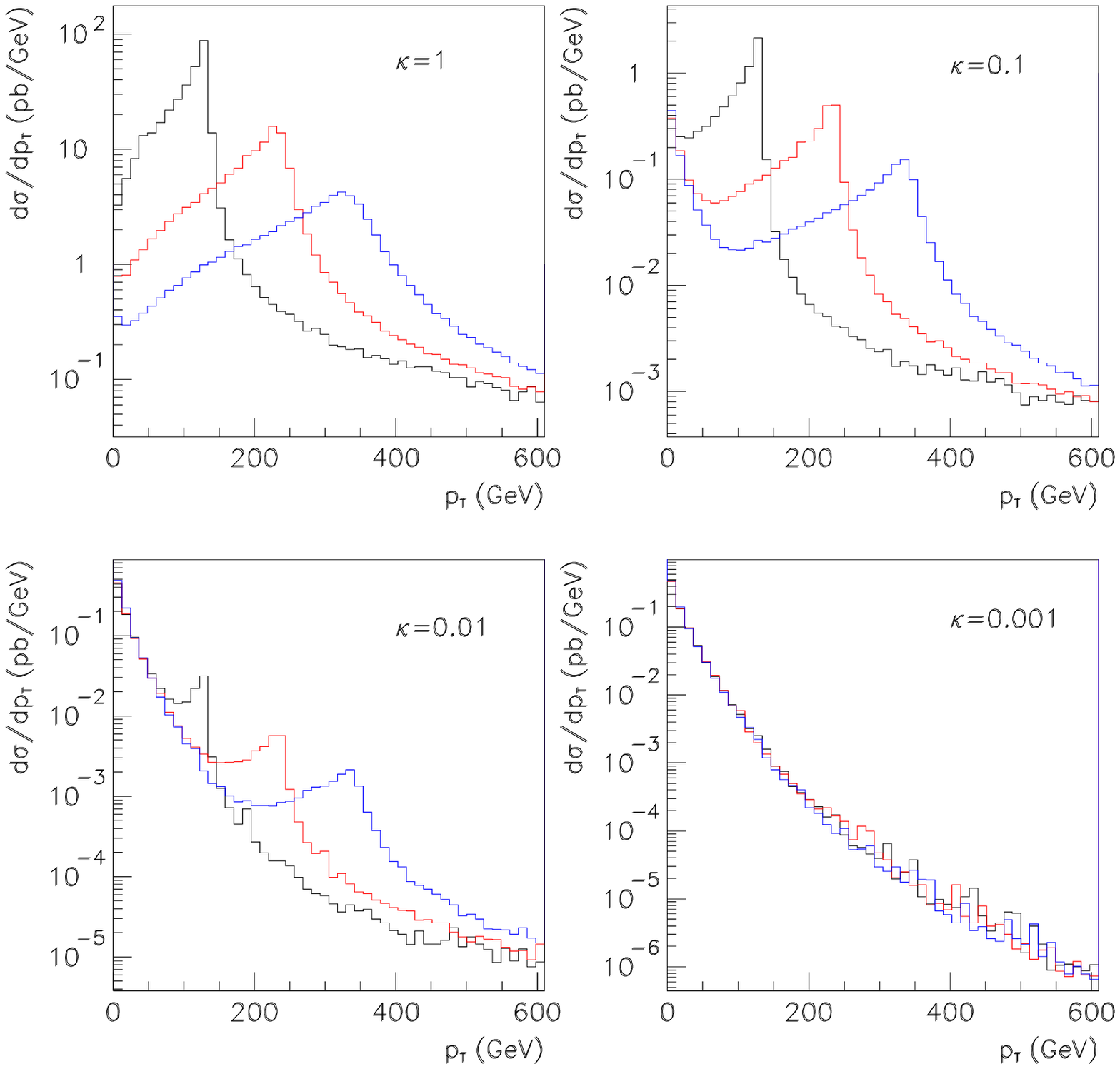}

\caption{Partonic level $p_{T}$ distributions of $b$ quark from the signal+background
process $pp\to W^{+}bX$ at $\sqrt{s}=14$ TeV. The peaks occur around
respective half mass values of the $t'$ quark masses.\label{fig:7}}

\end{figure}

\section{Signal and Background }

\begin{figure}
\includegraphics[scale=0.4]{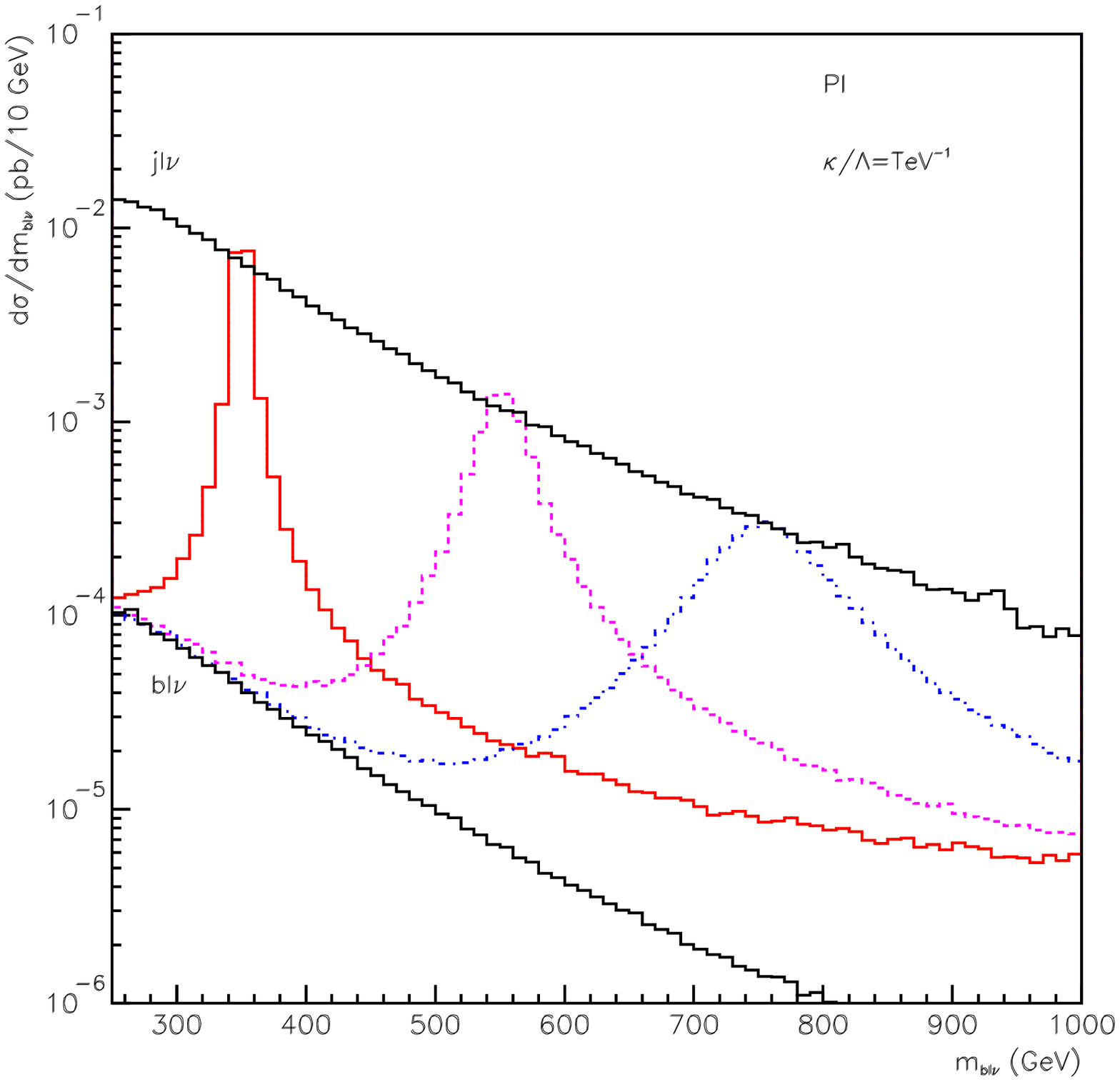}\includegraphics[scale=0.4]{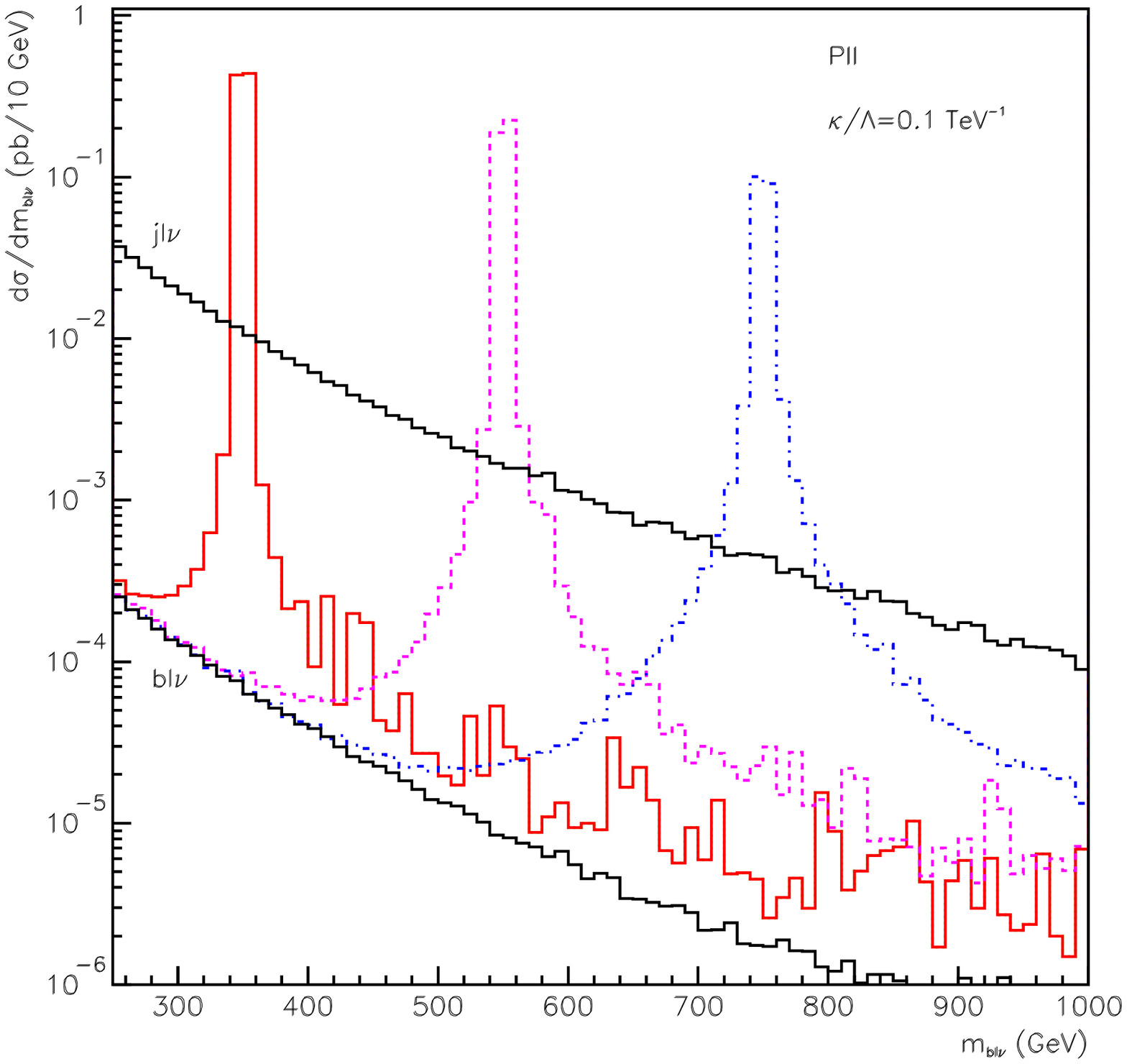}

\includegraphics[scale=0.4]{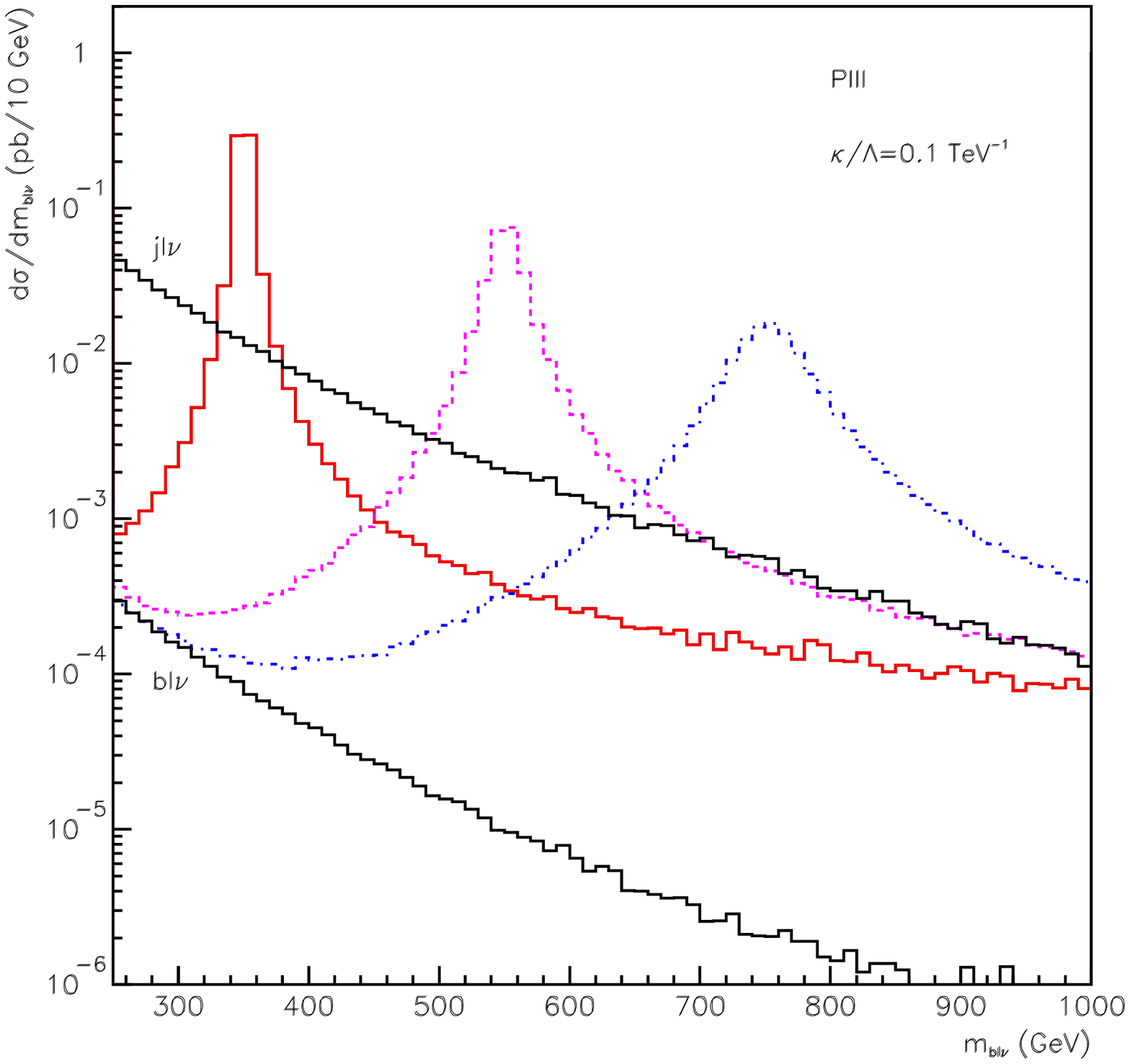}

\caption{Invariant mass distributions of $b\ell\nu$ system at partonic level
for PI (with $p_{T}^{b,j}>100$ GeV), PII (with $p_{T}^{b,j}>50$
GeV), and PIII (with $p_{T}^{b,j}>50$ GeV) parametrizations at $\sqrt{s}=14$
TeV.\label{fig:8}}

\end{figure}

\begin{figure}
\includegraphics[scale=0.8]{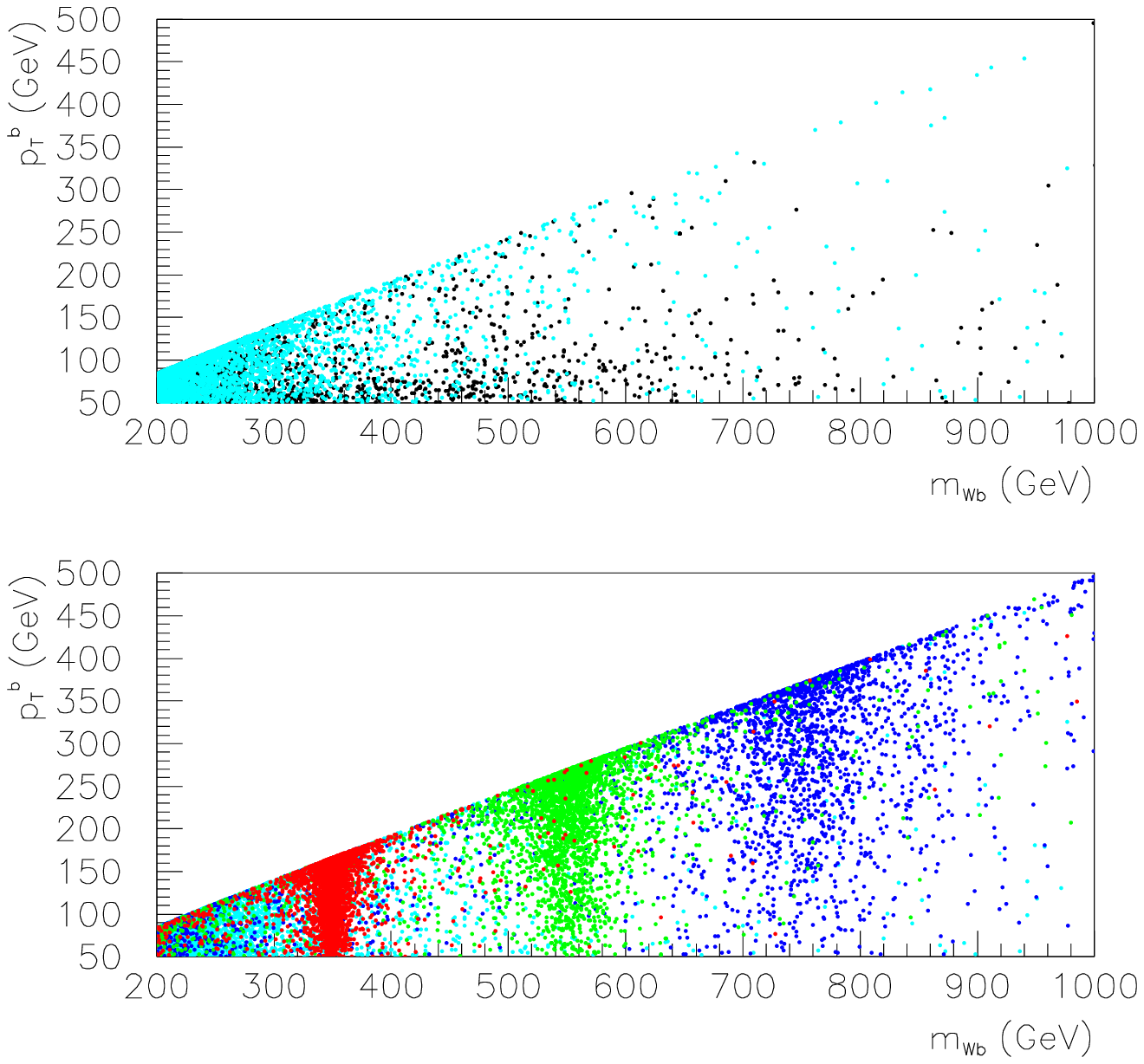}

\caption{Transverse momentum of the $b$-jet versus $W^{+}b$ invariant mass
at $\sqrt{s}=14$ TeV, on the upper panel only the background distribution
with $W^{+}b$ (black) and $W^{+}j(b)$ (cyan) events is shown, while
on the lower panel both the background and signal distribution with
mass of $350$ (red), $550$ (green) and $750$ GeV (blue). Here,
the anomalous coupling is taken to be $\kappa/\Lambda=1$ TeV$^{-1}$
and the mixing with the fourth family quark $t'$ is parametrized
as the PII.\label{fig:9}}
 
\end{figure}

The resonant production mechanisms of the fourth family $t'$ quarks
depends on the anomalous coupling $\kappa/\Lambda$, while their anomalous
decays and charged current decays depend on both these couplings and
the $4\times4$ CKM matrix elements. The signal process $pp\to W^{+}bX$
includes $t'$ exchange in the $s$-channel. The $s-$channel contribution
to the signal process would manifest itself as resonance around the
$t'$ mass value in the $W$-boson+jet reconstructed invariant mass.
When we consider the leptonic $W$-decays, the $t'$ signal search
will be $\ell^{+}+b_{jet}+\not E_{T}$, where $\ell=e,\mu$. For the
hadronic $W$-decays one would seek the events with one $b$-jet alongside
two more jets requiring these to have an invariant mass peak around
the $W$-mass. If we consider the dominance of the SM decay mode over
the anomalous decay, the $t'$ resonant production signal will be 

\begin{equation}
\begin{array}{cccc}
pp\to & t'(\bar{t}') & X\\
 & \hookrightarrow & W^{+(-)} & b_{jet}\\
 & \hookrightarrow & W^{+(-)} & j\end{array}\label{eq:13}\end{equation}
which includes further leptonic or hadronic decays of $W^{+}$boson.
In the second and third lines the elements of $4\times4$ CKM matrix
$V_{t'b}$ and $V_{t'q}$ enter to the decay process for $t'$ signal.
The cross section for the SM process $pp\to W^{+}bX$ ($pp\to W^{-}\bar{b}X$)
is 10.14 pb (9.78 pb) without any cuts at $14$ TeV, and 5.73 pb (5.49
pb) at $10$ TeV. For the cross section estimates, we assume the efficiency
for $b$-tagging as $\epsilon_{b}=50\%$, and rejection factors $r_{j}=100$
for light jets, and $r_{c}=10$ for $c(\bar{c})$ quark jets since
they are assumed to be mis-tagged as $b$-jets. The $p_{T}$ distributions
for both signal and background are given in Fig.\ref{fig:7} including
the interference terms. Moreover, different background processes contributing
to the same final state are presented in Table \ref{tab:tab6} with
various $p_{T}$ cuts. At the center of mass energy $\sqrt{s}=10$
TeV, the background cross sections are calculated as $0.375$ pb (0.354
pb) for $W^{+}b$ ($W^{-}b$), and $2.69\times10^{3}$ pb (1.88$\times10^{3}$
pb) for $W^{+}j$ ($W^{-}j$) with $p_{T}^{j}>50$ GeV. The invariant
mass distributions of $b\ell\nu$ system at partonic level for different
parametrizations are given in Fig. \ref{fig:8}. Here, the $W^{+}b$
and $W^{+}j$ backgrounds are included with the assumed efficiencies
and acceptance factors. In Fig. \ref{fig:9}, we show a density plot
of the transverse momentum of the $b$ quark and the invariant mass
distribution of $W^{+}b$ system with the $t'$ mass constraint.

\begin{table}
\caption{Cross sections for the backgrounds ($W^{\pm}b$ and $W^{\pm}j$) with
$p_{T}$ cuts at the center of mass energy $\sqrt{s}=14$ TeV.\label{tab:tab6}}

\begin{tabular}{|c|c|c|c|}
\hline 
Backgrounds & $p_{T}>20$GeV & $p_{T}>50$GeV & $p_{T}>100$GeV\tabularnewline
\hline 
$W^{+}b(W^{-}\overline{b})$ & $2.79$($2.71$) & $6.98\times10^{-1}$($6.71\times10^{-1}$) & $1.16\times10^{-1}$($1.09\times10^{-1}$)\tabularnewline
\hline 
$W^{+}j(W^{-}j)$ & $2.22\times10^{4}$($1.64\times10^{4}$) & $5.37\times10^{3}$($3.87\times10^{3}$) & $1.02\times10^{3}$($6.92\times10^{2}$)\tabularnewline
\hline
\end{tabular}
\end{table}

\section{Analysis}

At the generator level, we have required a $b$-jet with transverse
momentum at least $p_{T}^{b}>50$ GeV for the $W^{+}b$ events. The
events generated for each subprocess are mixed using the \emph{{}``mix''}
script which can be found in the CompHEP package \cite{comphep},
and passed to the PYTHIA \cite{pythia64} for further decays and hadronization
using the cpyth package \cite{belyaev00}. After $W$-boson decay
and hadronization, the detector effects, such as acceptance and resolution
are simulated with PGS4 program \cite{conley} using generic LHC detector
\cite{ATLAS99} parameters. This fast simulation includes the most
important detector effects, such as $p_{T}$ smearing, $E_{em}$ and
$E_{had}$ smearing, energy deposited in towers (granularity) and
tag efficiencies of the remaining detector effects such as mis-identifications.
More realistic simulation requires the resources of the LHC Collaborations
which are beyond the scope of this work. ExRootAnalysis package \cite{ExRoot}
is used to PGS4 data and the output is analysed and histogrammed with
the ROOT \cite{ROOT} macros. Since the cross section for $t'\to W^{+}b$
is about five times larger than $\bar{t'}\to W^{-}\bar{b}$ cross
section, the analysis for the former process has been considered for
the remaining of this study.

\begin{figure}
\includegraphics[scale=0.4]{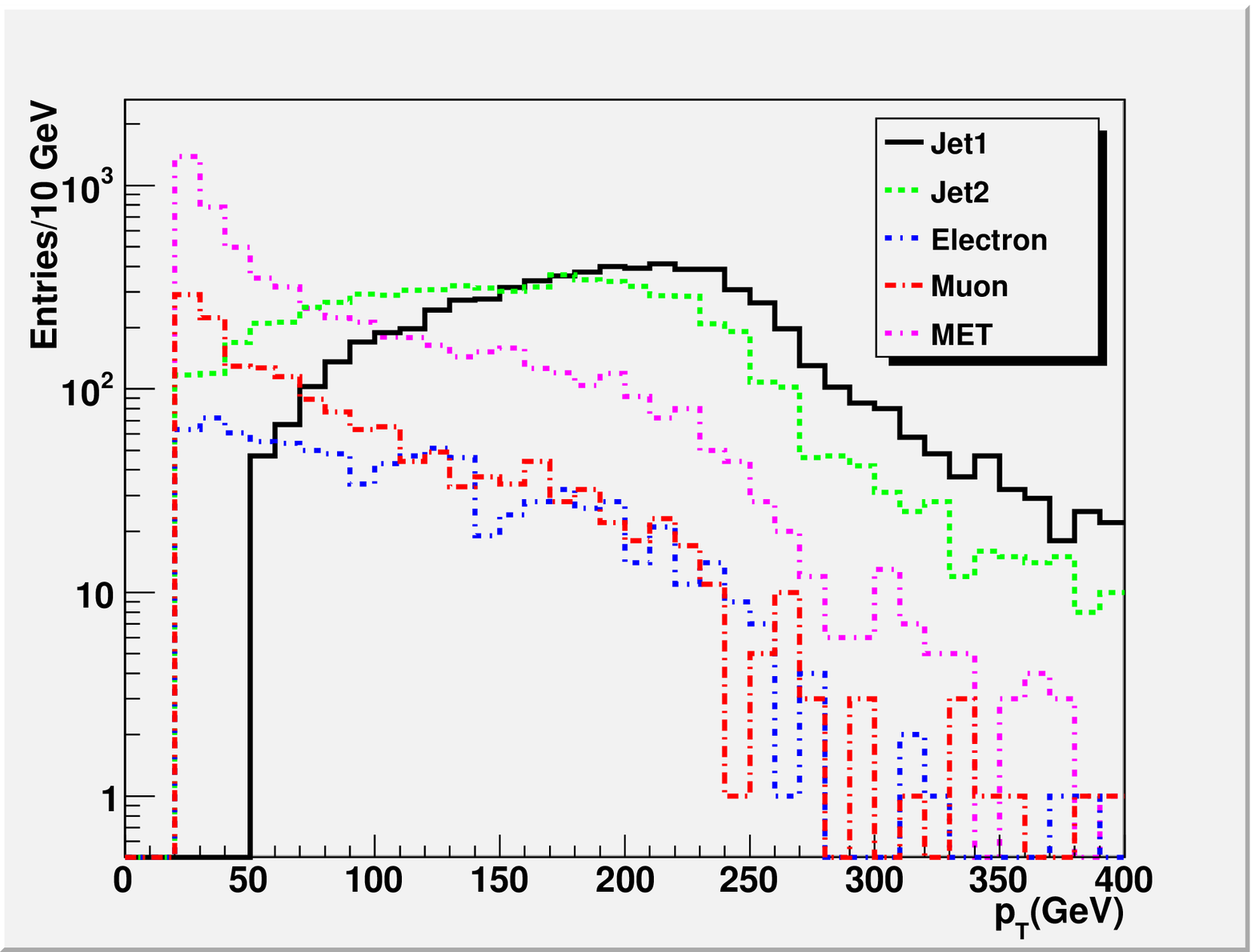}\includegraphics[scale=0.4]{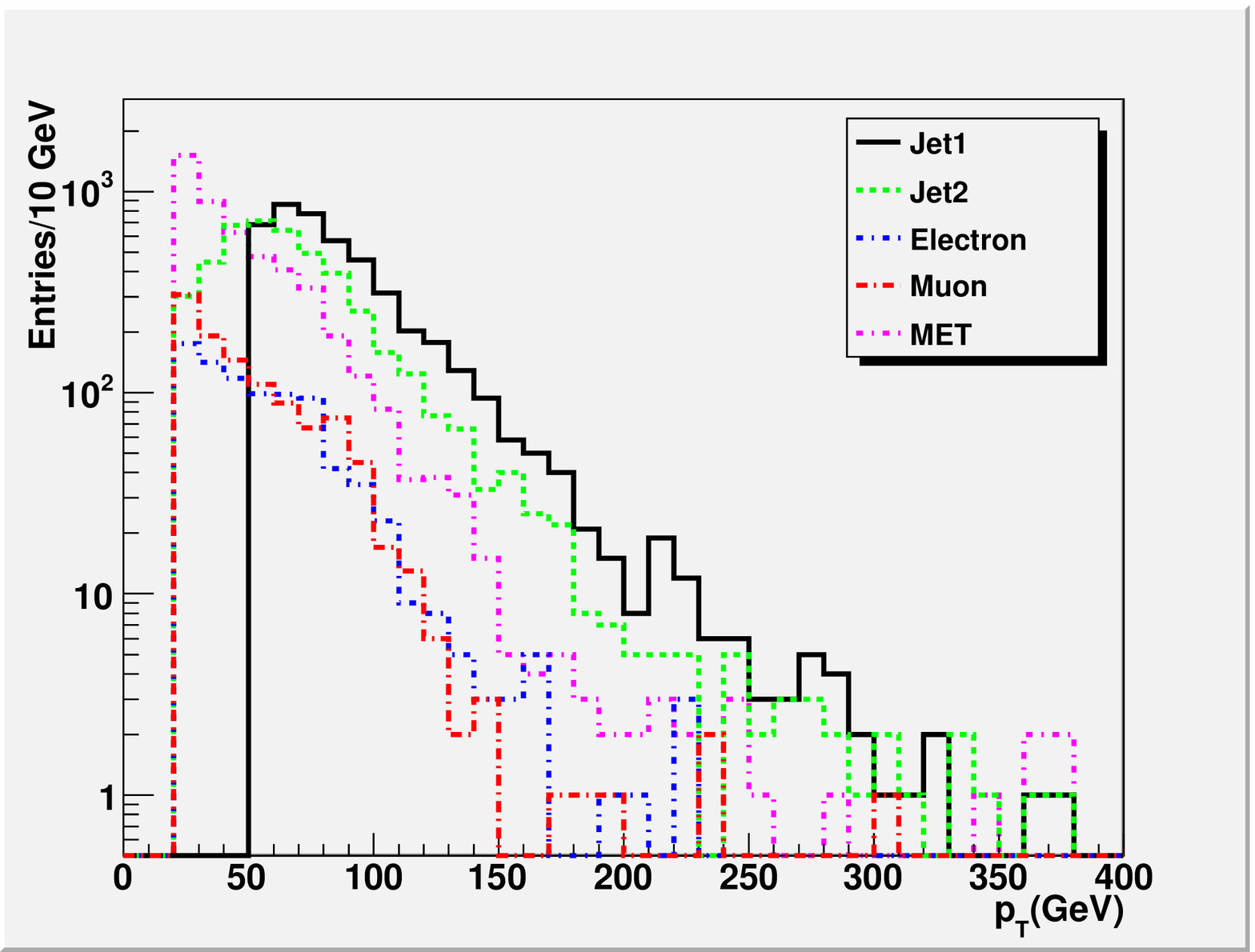}

\caption{The $p_{T}$ distributions of jets, electrons and muons from the signal
($t'\to W^{+}b$) on the left panel, and background ($W^{+}b$) on
the right panel shown after detector simulation. The jets are ordered
according to the magnitude of their transverse momenta, here we apply
the cuts $p_{T}^{j_{1}}>50$ GeV, $p_{T}^{j_{2,3}}>20$ GeV, $p_{T}^{\ell}>20$GeV
and $\not\! E_{T}>20$ GeV. \label{fig:10}}

\end{figure}

\begin{figure}
\includegraphics[scale=0.4]{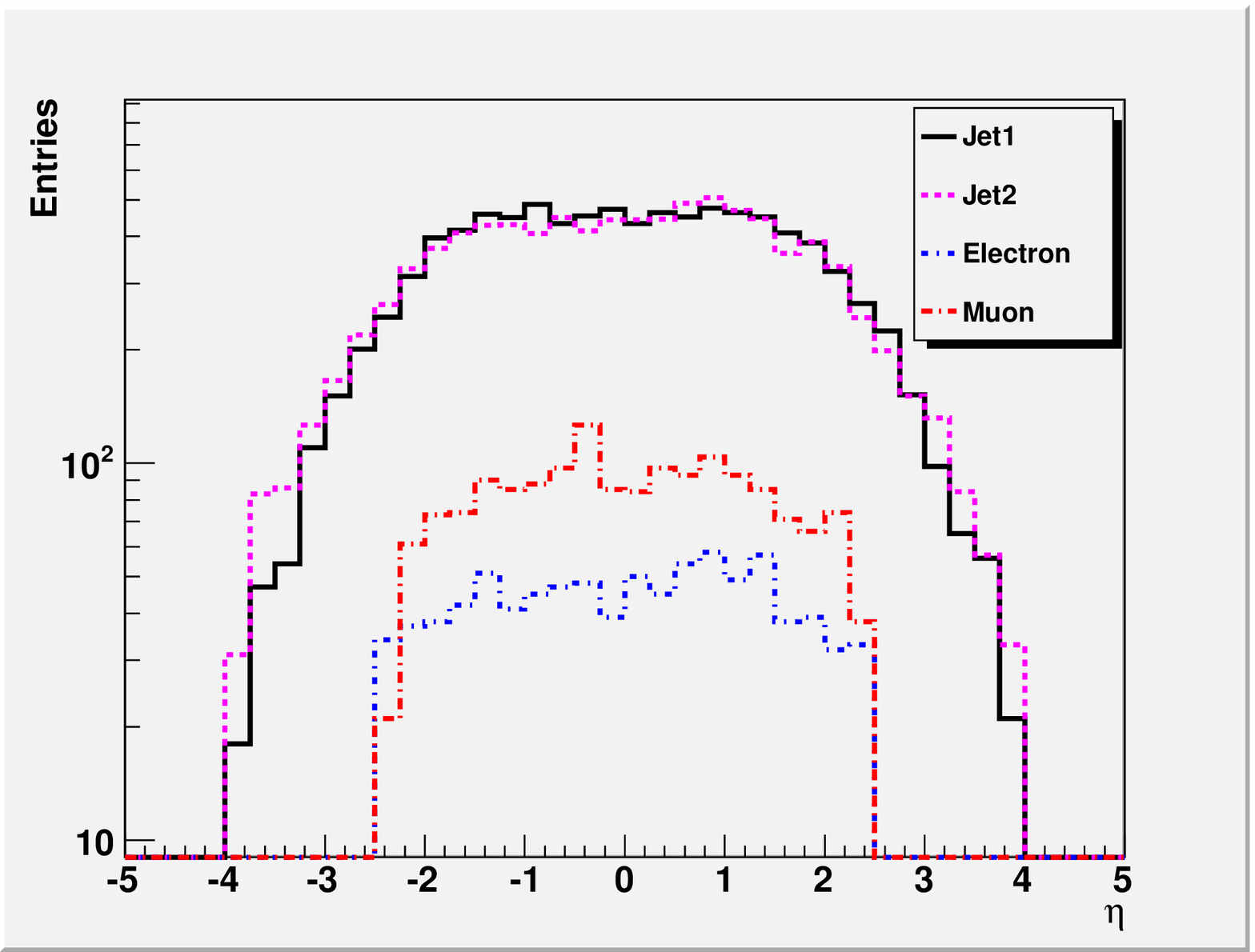}\includegraphics[scale=0.4]{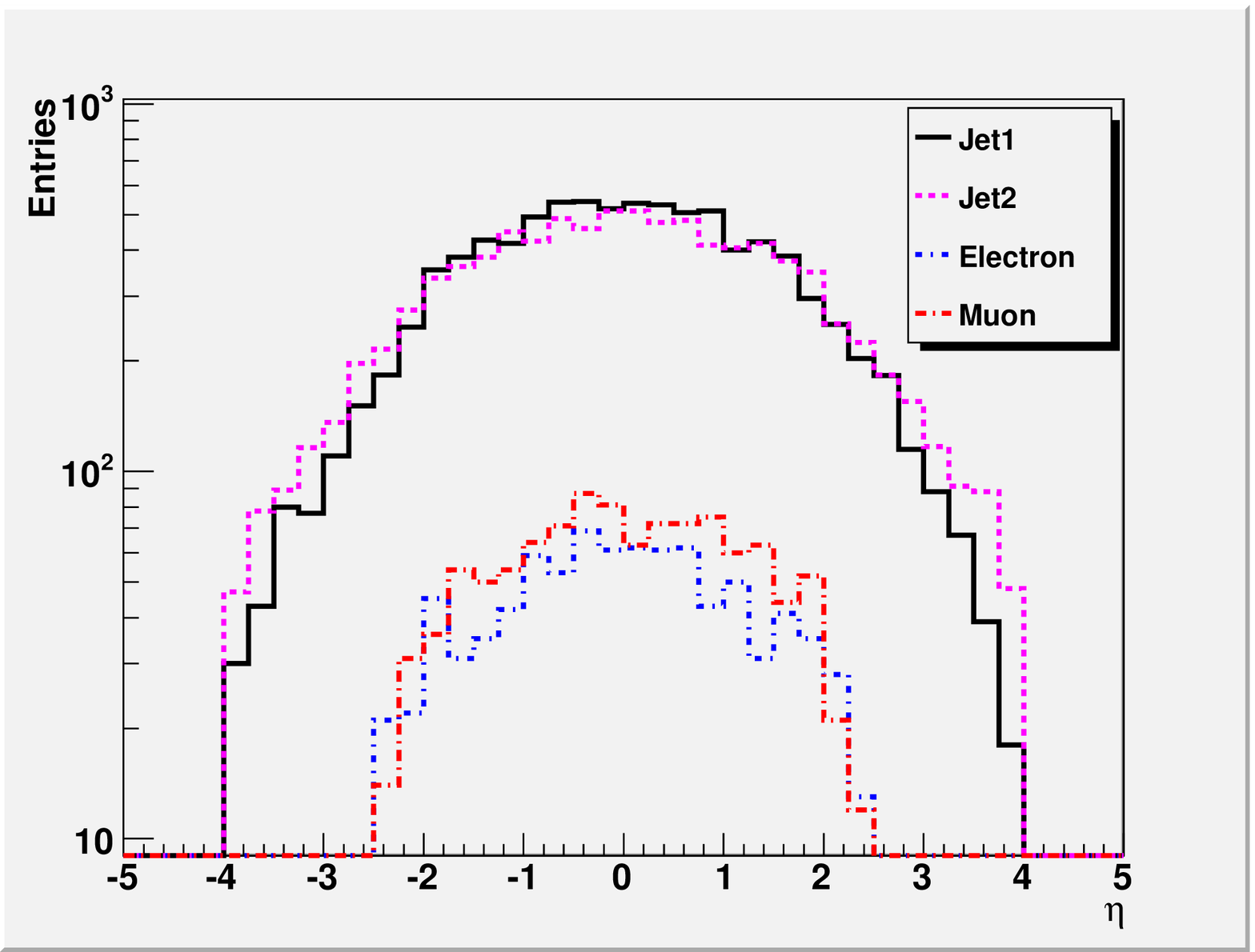}

\caption{The $\eta$ distributions of jets, electrons, muons from the simulated
signal ($t'\to W^{+}b$) on the left panel, and background ($W^{+}b$)
on the right panel. The jets are ordered according to the magnitude
of their $p_{T}$. Typical LHC detectors have the acceptance for $|\eta|<2.5$
when utilization of inner tracker is imposed.\label{fig:11}}

\end{figure}

Typical kinematical distributions are shown in Figs.\ref{fig:10}
and \ref{fig:11}. In the event analysis, the signal ($t'\to W^{+}b$,
with $\kappa/\Lambda=0.1$ TeV$^{-1}$ and $m_{t'}=400,500,600$ GeV)
and background ($W^{+}b$) are taken into account assuming PIII parametrization.
The $W$-boson invariant mass can be reconstructed from its leptonic
or hadronic decays. For the leptonic reconstruction case, the criteria
is applied to the electrons or muons are: $p_{T}^{\ell}>20$ GeV and
$|\eta^{\ell}|<2.5$ whereas for missing transverse energy, the requirement
is $\not\! E_{T}>20$ GeV. For the hadronic reconstruction case, we
require to find at least 2 jets with $p_{T}^{j}>20$ GeV and $|\eta^{j}|<2.5$.
For the reconstruction of the $t'$ quark invariant mass, the $b$-tagged
jets are required to have $p_{T}^{j_{b}}>50$ GeV and $|\eta^{j_{b}}|<2$
for both cases.

\begin{figure}
\includegraphics[scale=0.5]{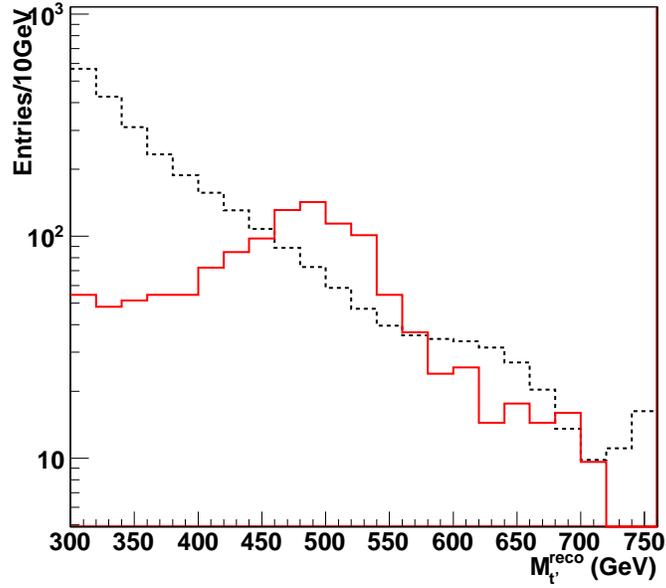}

\caption{Invariant mass distribution of $b\ell\nu$ system for both signal
($\kappa/\Lambda=0.1$ TeV$^{-1}$) and background at $\sqrt{s}=14$
TeV.\label{fig:12}}

\end{figure}

\begin{figure}
\includegraphics[scale=0.4]{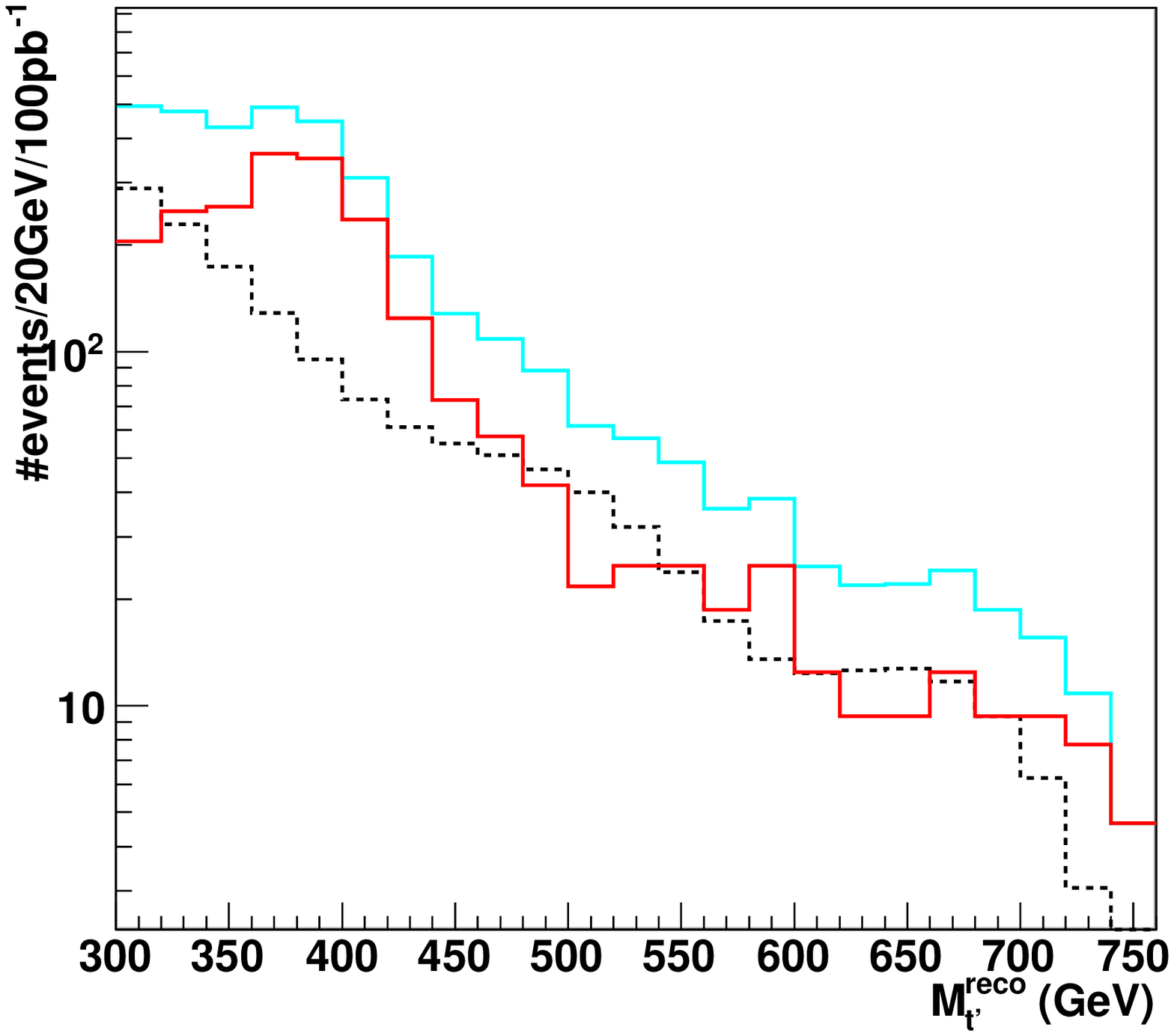}\includegraphics[scale=0.4]{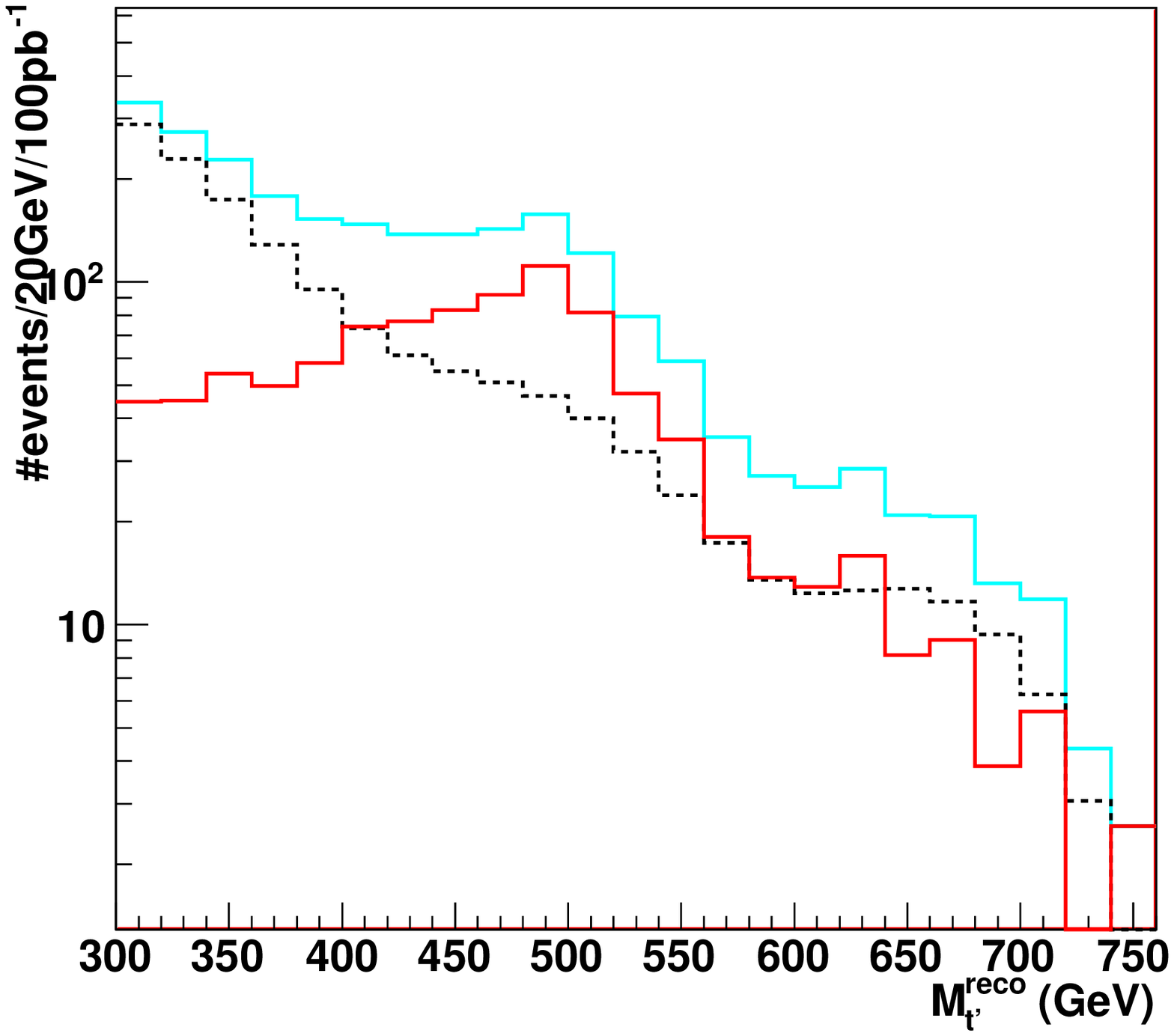}

\includegraphics[scale=0.4]{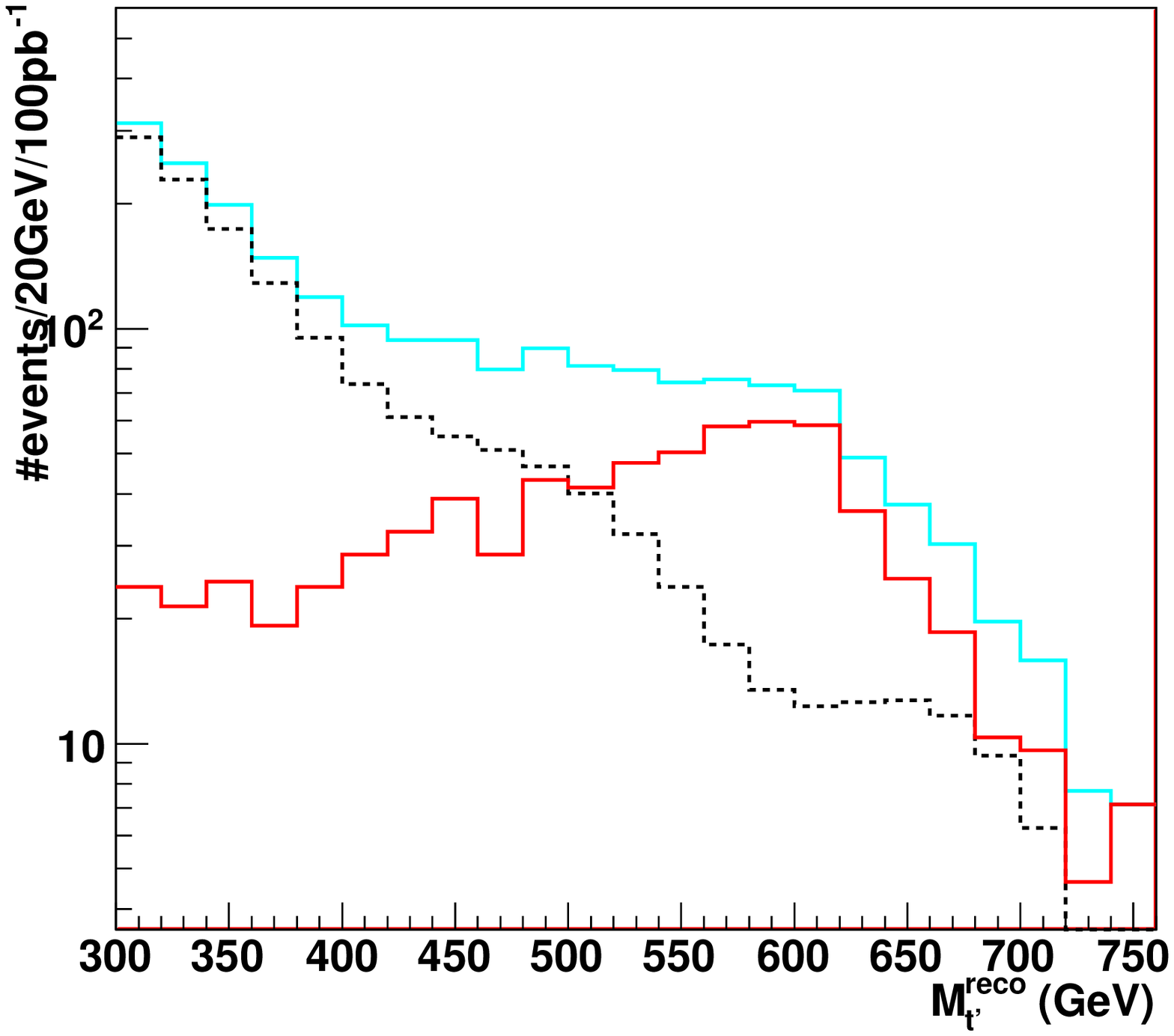}

\caption{Mass reconstruction of the $t'$ signal ($b\ell\nu$) (solid) with
$\kappa/\Lambda=0.4$ TeV$^{-1}$ for $m_{t'}=400$ GeV (upper-left),
$m_{t'}=500$ GeV (upper-right) and $m_{t'}=600$ GeV (lower), and
the corresponding background (dashed) at $\sqrt{s}=10$ TeV.\label{fig:13}}

\end{figure}

The final results are presented only for the case where the $W$ boson
is reconstructed from its leptonic decays. The four momentum vector
of the neutrino is calculated from lepton and missing transverse energy
assuming a $W$ boson rest mass constraint. The plots for the reconstucted
$t'$ invariant mass after detector simulation is given in Fig.\ref{fig:12}
at $\sqrt{s}=14$ TeV and in Fig.\ref{fig:13} at $\sqrt{s}=10$ TeV.
We include the possible backgrounds contributing to the same final
state, and count the signal ($S$) and background ($B$) events in
the corresponding mass intervals to calculate the statistical significance
(\emph{SS}) defined as \cite{CMSTDR} :

\begin{equation}
SS=\sqrt{2\left[(S+B)\ln(1+\frac{S}{B})-S\right]}\label{eq:15}\end{equation}
In Table \ref{tab:7}, the significance calculations are presented
for different mass and anomalous coupling values at $\sqrt{s}=10$
TeV. Here, we use the mass bin width $\Delta m=\max(2\Gamma,\delta m)$
to count signal and background events with the mass resolution $\delta m$.
The significance increases with $\kappa/\Lambda$ assuming a maximal
mixing between the fourth and the third family quarks. The results
of this study show that, with early LHC data, one can discover extra
up-type quark if there is large anomalous coupling with other up-type
quarks.

\begin{table}
\caption{The statistical significance ($SS$) for different masses of $t'$
quark, where we take some anomalous parameters in the range ($\kappa=0.1-1.0$)
at $\sqrt{s}=10$ TeV and $L_{int}=100$ pb$^{-1}$. \label{tab:7}}

\begin{tabular}{|c|c|c|c|}
\hline 
$SS$ & $m_{t'}=400$ GeV & $m_{t'}=500$ GeV & $m_{t'}=600$ GeV\tabularnewline
\hline
\hline 
$\kappa/\Lambda=0.1$ TeV$^{-1}$ & 1.91 & 1.99 & 2.25\tabularnewline
\hline 
$\kappa/\Lambda=0.2$ TeV$^{-1}$ & 6.95 & 7.19 & 8.11\tabularnewline
\hline 
$\kappa/\Lambda=0.4$ TeV$^{-1}$ & 21.09 & 21.40 & 23.44\tabularnewline
\hline 
$\kappa/\Lambda=0.6$ TeV$^{-1}$ & 41.97 & 38.51 & 38.28\tabularnewline
\hline 
$\kappa/\Lambda=0.8$ TeV$^{-1}$ & 55.05 & 53.05 & 50.10\tabularnewline
\hline 
$\kappa/\Lambda=1$ TeV$^{-1}$ & 65.22 & 63.14 & 59.00\tabularnewline
\hline
\end{tabular}
\end{table}

\section{Conclusion}

Anomalous interactions could become significant at tree level processes
due to possible large mass of the fourth family quarks. The fourth
family $t'$ quarks can be produced with large numbers if they have
anomalous couplings dominates over the SM chiral interactions. Following
the results from the signal significance for $t'$ anomalous production
the sensitivity to anomalous coupling $\kappa/\Lambda$ can be reached
down to $0.1$ TeV$^{-1}$in the $b$-jet+lepton+$\not\! E_{T}$ channel
at $\sqrt{s}=10$ TeV assuming a maximal parametrization for the extended
CKM elements. The LHC experiments can observe the fourth family quarks
mostly in pairs and single in the $s$-channel if they have large
anomalous couplings to the known quarks. If detected at the LHC experiments
the fourth family quarks will change our perpective on the flavor
and the mass.
\begin{acknowledgments}
We acknowledge the support from CERN Physics Department. O.C. and
H.D.Y.'s work is supported by Turkish Atomic Energy Authority (TAEA)
and Turkish State Planning Organization under the grant no. DPT2006K-120470.
H.D.Y's work is also supported by TÜB\.{I}TAK with the project number
105T442. G.U.'s work is supported in part by U.S. Department of Energy
Grant DE FG0291ER40679. \end{acknowledgments}

\end{document}